\documentclass[a4paper,11pt]{article}
\usepackage{jcappub} % for details on the use of the package, please see the JINST-author-manual
\usepackage{lineno}
\newcommand{\II}{\mathrm i}

\title{\boldmath Study of a cubic cavity resonator for gravitational waves detection in the microwave frequency range}

\author{Pablo Navarro,$^a$ Benito Gimeno,$^b$ Juan Monz\'o-Cabrera,$^a$ Alejandro D\'iaz-Morcillo,$^a$ Diego Blas$^{c,d}$}
\affiliation{$^a$Departamento de Tecnolog\'ias de la Informaci\'on y las Comunicaciones,
Universidad Polit\'ecnica de Cartagena,
Plaza del Hospital 1, 30302 Cartagena, Spain,\\
$^b$Instituto de Física Corpuscular (IFIC), CSIC-University of Valencia, Calle Catedr\'atico Jos\'e Beltr\'an Martínez, 2, 46980 Paterna (Valencia), Spain, \\
$^c$Institut de F\'isica d’Altes Energies (IFAE), The Barcelona Institute of Science and Technology, Campus UAB, 08193 Bellaterra (Barcelona), Spain\\
$^d$Instituci\'o Catalana de Recerca i Estudis Avan\c cats (ICREA), Passeig Llu\'is Companys 23, 08010 Barcelona, Spain}

\emailAdd{pablonm.ct.94@gmail.com, benito.gimeno@uv.es, juan.monzo@upct.es, alejandro.diaz@upct.es, dblas@ifae.es}

\abstract{The direct detection of gravitational waves (GWs) of frequencies above  MHz has recently received considerable attention. In this work we present a precise study of the reach of a cubic cavity resonator to GWs in the microwave range, using for the first time tools allowing to perform realistic simulations. Concretely, the BI-RME 3D method, which allows us to obtain not only the detected power but also the detected voltage (magnitude and phase), is used here. After analyzing three cubic cavities for different frequencies and working simultaneously with three different degenerate modes at each cavity, we conclude that the sensitivity of the experiment is strongly dependent on the polarization and incidence angle of the GW. The presented experiment can reach sensitivities up to  $ 1 \cdot 10^{-19}$ at 100\, MHz, $ 2 \cdot 10^{-20}$ at 1\, GHz, and $ 6 \cdot 10^{-19}$ at 10\, GHz for optimal angles and polarizations, and where in all cases we assumed an integration time of  $\Delta t = 1$ ms. These results provide a strong case for further developing the use of cavities to detect GWs. Moreover, the possibility of analyzing the detected voltage (magnitude and phase) opens a new interferometric detection scheme based on the combination of the detected signals from multiple cavities.}

\begin{document}

\maketitle
\flushbottom

\section{Introduction}
\label{sec:introduction}

\noindent 
The direct detection of gravitational waves (GWs) by Earth interferometers opened a new era in our quest to understand the constituents of the Universe~\cite{LIGOScientific:2016aoc}. These detectors have been able to explore GWs in a band around $10^2$ Hz~\cite{Maggiore}. Equally relevant are the recent detections by pulsar timing arrays (PTA) \cite{Antoniadis:2023rey,Reardon:2023gzh,NANOGrav:2023gor,Xu:2023wog} of GWs in the nHz band, whose origin is still to be determined. 
 The exploration of GWs in other frequency bands is a very active and promising field of research. This is partly due to the rich variety of sources expected in basically all the band from $10^{-17}$ Hz (covered by CMB observations~\cite{Namikawa:2019tax}) to 10 kHz. 
 Of particular interest for future searches are the space-based interferometers, such as LISA~\cite{LISA:2017pwj}, with maximum sensitivity around $10^{-3}$\,Hz and an impressive scientific legacy. Ideas abound to continue exploring GWs in the band from nHz to kHz, such as updated ground-based laser interferometers~\cite{Hild:2010id, Punturo:2010zz,LIGOScientific:2016wof}, atom interferometers~\cite{Badurina:2019hst,MAGIS-100:2021etm,AEDGE:2019nxb}, other 
space missions \cite{Yagi:2011wg,Sesana:2019vho}, or proposals related to orbital tracking~\cite{Blas:2021mqw,Fedderke:2021kuy}. As mentioned, this fantastic coverage is particularly relevant given the number of signals
expected to be present, carrying information about astrophysics as well as fundamental physics.

The extension to frequencies above 10 kHz has been less explored for different reasons.  First, as the frequency $f$ grows, the devices most sensitive to these GWs of shorter wavelength become smaller and may also require faster readouts. Furthermore, and relevant for interferometers, the effect of GWs on the relative distance of test masses at distances $L$ decreases as $\delta L\propto  h\, L$, with $h$ being the amplitude of the wave. Finally,  it is not clear which astrophysical/fundamental sources may produce GWs at $f\gtrsim 10$\, kHz at levels that will impact our detectors. It has been recently emphasized that these aspects may be turned into new opportunities \cite{Aggarwal:2020olq}: on the one hand, smaller devices may imply a connection with the cutting-edge precision of the most sensitive sensors, and call for rethinking the possible detection strategies. On the other hand, the absence of astrophysical sources may be also positive, since any {significant} signal may mean a deviation from the standard model. In fact, there are {known} candidates for GWs at high frequencies of stochastic or coherent nature, though the expectation is that any detection by current technologies may point towards new physics~\cite{Aggarwal:2020olq}. Independently of these points,  
being at the dawn of the exploration of high-frequency gravitational waves (HFGWs), it is almost unquestionable the relevance of understanding its current limits and future prospects.

Out of the possible effects that a GW may generate as it traverses a laboratory, we will focus on those related to the current generated by a GW in the presence of a background electromagnetic field,  called {the inverse Gertsenshtein} effect when the electromagnetic background field is a static magnetic field \cite{gertsenshtein1962wave,gertsenshtein_1962} (see also \cite{Berlin:2023grv,Domcke:2023qle} for other recent proposals to detect HFGWs).
As a result,  {any} technique used to detect ({low energy}) photons in the presence of an electromagnetic background may, in principle, be used as a GW detector.  This is the idea behind the recent proposals, such as \cite{Herman:2020wao,Berlin:2021txa,Domcke:2023bat, Domcke:2023qle}, which consider experiments conceived to detect the conversion of {axions} into photons in the presence of a magnetic field as GW detectors. Our goal in this paper is to perform a study of this effect with the advanced tools and language used in the simulations of radio frequency resonant cavities.  This is the first step towards the more ambitious goal of fully characterizing the optimal way to search for HFGWs in these devices and start exploring new read-out ideas before performing dedicated searches (see the recent  \cite{Domcke:2023bat} for related work in this direction).

In this work, we apply a full-wave modal technique for the rigorous electromagnetic study of the coupling GWs-cavity which is based on the advanced modal technique Boundary Integral - Resonant Mode Expansion (BI-RME) 3D \cite{conciauro}. This method allows for an efficient characterization of the interaction between the GW and the resonant cavity. The GWs generate an externally induced current which excites the resonant modes of the cavity under the presence of an intense static magnetic field. Using the BI-RME 3D formulation, we can represent the excited resonator in terms of an equivalent network driven by the current sourced by GWs. Finally, we will compute both the voltage and the power extracted from the cavity obtaining information about the magnitude and the phase of the detected signal. As an application, we have designed a cubic resonator with three orthogonal coaxial antennas which allow for the synchronous detection of an incident GW; both polarizations of the GWs have been accounted for.

This paper is organized as follows. After this introduction, we review the basic theory for the derivation of the current densities induced by the GWs in the presence of a stationary magnetic field in section~\ref{sec:gw-em}. Next, we introduce the BI-RME 3D theory in section~\ref{sec:bi-rme3d_formulation}. As an application of the presented formulation, in section~\ref{sec:application} we design three cubic resonators tuned at $100$ MHz, $1$ GHz, and $10$ GHz, obtaining sensitivities, and extracted power levels and voltages for the best sensitivities. Finally, conclusions and future research lines are presented in section~\ref{sec:conclusions}.

%%%%%%%%%%%%%%%%%%%%%%%%%%%%%%%%%%%%%%%%%%%%%%%%%%%
\section{The system GW-EM fields: induced current}
\label{sec:gw-em}
%%%%%%%%%%%%%%%%%%%%%%%%%%%%%%%%%%%%%%%%%%%%%%%%%%%

As shown in  \cite{Domcke:2022rgu} (see also \cite{Herman:2020wao,Berlin:2021txa}), in the presence of a GW with field values $h_{\mu\nu}(t,\vec r)$, $\Vec{r}$ being the position vector and $t$ the time measured in the reference frame of the laboratory, Maxwell's equations are modified to\footnote{Recall that Greek letters run as $\mu={0,1,2,3}$, while Latin letters have only spatial part $i={1,2,3}$. We   use the $\eta_{\mu\nu}={\rm diag}(-+++)$ metric convention. In this section, we will stick to the units $c=1$. For the calculations in the following sections, we will convert to the SI. Two repeated indexes are always summed over.}

\begin{equation}
    \partial_\nu F^{\mu \nu}=j_{\mathrm{eff}}^\mu=\left(-\nabla \cdot \vec{P}, \nabla \times \vec{\tilde M}+\partial_t \Vec{P} \right),\label{eq:max_mod}
\end{equation}
where $j_{\mathrm{eff}}^\mu$ is the effective induced current density and
\begin{equation}
    \begin{aligned}
P_i & =-h_{i j} E_j+\frac{1}{2} h E_i+h_{00} E_i-\epsilon_{i j k} h_{0 j} B_k ,\\
\tilde M_i & =-h_{i j} B_j-\frac{1}{2} h B_i+h_{j j} B_i+\epsilon_{i j k} h_{0 j} E_k,
\end{aligned}
\end{equation}
where $h=\eta^{\mu\nu}h_{\mu\nu}$ is the trace of the GW, $\epsilon_{ijk}$ is the Levi-Civita symbol, $E_i$ are the components of the electric field, and $B_i$ those of the magnetic field. Quite relevantly, the values of $h_{\mu\nu}$ for a GW seen in the laboratory are those associated with the so-called {laboratory frame}, which for a GW of angular frequency $\omega = 2 \pi f$ propagating in the $\Vec{k}$ direction ($\Vec{k} =\omega \, \hat{k}$) read  \cite{Berlin:2021txa,Domcke:2022rgu}
\begin{equation}
\begin{aligned}
& h_{00}=\omega^2 F(\vec{k} \cdot \vec{r}) \, \vec{b} \cdot \vec{r},\left.\quad b_j \equiv r_i h_{i j}^{\mathrm{TT}}\right|_{\vec{r}=\Vec{0}}, \\
& h_{0 i}=\frac{1}{2} \omega^2\left[F(\vec{k} \cdot \vec{r})+\mathrm{i} F^{\prime}(\vec{k} \cdot \vec{r})\right]\left(\hat{k} \cdot \vec{r} \, b_i \, - \, \vec{b} \cdot \vec{r} \, \hat{k}_i\right), \\
& h_{i j}=\mathrm{i} \omega^2 F^{\prime}(\vec{k} \cdot \vec{r})\left(\left.||\vec{r}||^2 h_{i j}^{\mathrm{TT}}\right|_{\vec{r}=\Vec{0}}+\vec{b} \cdot \vec{r} \delta_{i j} \, - \, b_i r_j \, - \, b_j r_i\right),
\end{aligned}
\end{equation}
 where $F(x)=\left(e^{-\mathrm{i} x}-1+\mathrm{i} x\right) / x^2$, $\hat{k}$ is the unitary vector in the propagation direction of the GW, and
 $\mathrm i = \sqrt{-1}$ is the imaginary unit. In the previous equation, $\left.h_{i j}^{\mathrm{TT}}\right|_{\Vec{r}=\vec{0}}$ represents the  GW as seen in the {transverse-traceless} coordinate system~\cite{Maggiore}, 
 \begin{equation}
     h_{i j}^{\mathrm{TT}}=\left[\left(\mathrm{U}_i \mathrm{U}_j-\mathrm{V}_i \mathrm{~V}_j\right) h^{+}+\left(\mathrm{U}_i \mathrm{~V}_j+\mathrm{V}_i \mathrm{U}_j\right) h^{\times}\right]\frac{e^{\II (\omega t-{\vec{k}} \cdot\vec{r})}}{\sqrt{2}},
     \label{eq:TT}
 \end{equation}
 evaluated at the origin $\Vec{r} = \Vec{0}$. Finally, the vectors generating the tensor structure for the polarizations $h_{+,\times}$ have to be orthogonal to $\hat{k}$ and of unit norm. In the cases we will explore, we will assume that the wave is traveling along one of the Cartesian planes, and $\mathrm{V}_i$ will be chosen as the Cartesian coordinate perpendicular to it while $\mathrm{U}_i=\epsilon_{ijk} \mathrm{V}_j \hat k_k$. From the previous expressions, one can now compute the effective current for different geometries. The final form of the current is not particularly illuminating, except for simple cases, and is omitted.   

%%%%%%%%%%%%%%%%%%%%%%%%%%%%%%%%%%%%%%%
\section{The BI-RME 3D formulation}
\label{sec:bi-rme3d_formulation}
%%%%%%%%%%%%%%%%%%%%%%%%%%%%%%%%%%%%%%%

In the second half of the last century, many research groups worldwide developed several numerical techniques for the electromagnetic analysis of microwave passive components and circuits. Some of them, such as the Finite Element Method (FEM) \cite{jin,salazar}, the Finite Different in Time Domain (FDTD) \cite{taflove}, and the Transmission Line Method (TLM) \cite{itoh}, have a very general range of applicability in complex geometries including the presence of dielectric and/or magnetic media, and, for these reasons constitute nowadays the basis of many commercial software \cite{CST,HFSS,COMSOL}. Besides, other numerical techniques such as the Method of Moments \cite{harrington} and the Mode Matching Method \cite{itoh} can deal only with specific components. However, these techniques require intense analytical processing, and, for this reason, they are not so popular, even though the developed codes are extremely fast and accurate \cite{hanson_yakovlev,chentotai_dyadic_green}.  

Among the modal methods proposed in the eighties and nineties, it should be mentioned the formulation proposed by Prof. Giuseppe Conciauro and co-workers at the Universit\`{a} degli Studio di Pavia (Italy) called the BI-RME 3D method, which
represents an advanced full-wave modal technique for the accurate and efficient electromagnetic analysis of microwave arbitrarily-shaped cavities \cite{conciauro,bi-rme3d_3Dcavities,bi-rme3d_3Dcavities_ports} including metallic obstacles \cite{bi-rme-3d_fermin,bi-rme-3d_angel_posts}. The complete formulation and the different implementations are very extensive and can be found in the technical literature \cite{bi-rme3d_overview}.

In the case of a microwave cavity resonator connected to different $P$ waveguide-ports, the starting point of the BI-RME 3D method is to express the time-harmonic (complex phasors) electric and magnetic fields generated by volumetric electric current density sources $\vec{J}$ and surface magnetic current density sources $\vec{M}$ within the cavity with the following integral expressions (given  now in the SI system),
\begin{eqnarray}   \label{bi-rme3d_E_general} 
	\vec{E}(\vec{r}) & = & \frac{\eta}{\mathrm{i} \, k} \, \nabla \int_V g^e
	(\vec{r},\vec{r}^{\, \prime}) \, \nabla' \cdot \vec{J}(\vec{r}^{\,
		\prime}) \, dV' - \mathrm{i} \, k \, \eta \int_V \mathbf{\vec{G}^{\rm A}}
	(\vec{r},\vec{r}^{\, \prime}) \cdot \vec{J}(\vec{r}^{\, \prime})
	\, dV' - \nonumber \\[0.3cm] & & - \int_S \nabla \times \mathbf{\vec{G}^{\rm
			F}} (\vec{r},\vec{r}^{\, \prime}) \cdot \vec{M}(\vec{r}^{\,
		\prime}) \, dS' \, + \, \frac{1}{2} \, \vec{n} \times \vec{M}   \,,
\end{eqnarray}	
and
\begin{eqnarray}  \label{bi-rme3d_H_general}
	\vec{H}(\vec{r}) & = & \frac{1}{\mathrm{i} \, k \, \eta} \nabla_s \int_S g^m
	(\vec{r},\vec{r}^{\, \prime}) \, \nabla' \cdot \vec{M}(\vec{r}^{\,
		\prime}) \, dS' + \frac{k}{\mathrm{i} \, \eta} \int_S \mathbf{\vec{G}^{\rm F}}
	(\vec{r},\vec{r}^{\, \prime}) \cdot \vec{M} (\vec{r}^{\, \prime})
	\, dS' + \nonumber \\[0.25cm] & & + \int_V \nabla \times \mathbf{\vec{G}^{\rm A}}
	(\vec{r},\vec{r}^{\, \prime}) \cdot \vec{J}(\vec{r}^{\, \prime})   
	\, dV'\,,
\end{eqnarray}
where $V$ is the simply connected volume of the empty resonator, which is bounded by perfectly conducting walls (conducting walls of the structure are lossless, but lossy walls will be further accounted for with the conventional perturbative method in equation~\eqref{eq:losses}). Vacuum is characterized by the electric permittivity $\varepsilon_0$ and the magnetic permeability $\mu_0$ of free space (dielectric and magnetic media can be accounted for in the BI-RME 3D theory, but they will not be considered in this work). In these equations $\eta = \sqrt{\mu_0/\varepsilon_0} \approx 120 \pi \, \Omega$ is the vacuum impedance; $k=\omega/c$ is the free space wavenumber; $c=1/\sqrt{\mu_0 \, \varepsilon_0}$ is the speed of light in vacuum; $\vec{n}$ is the inward unitary normal vector to the cavity surface; $\nabla_s$ is the surface divergence operator \cite{chentotai_vector_dyadic}; $g^e(\vec{r},\vec{r}^{\, \prime})$ and $g^m(\vec{r},\vec{r}^{\, \prime})$ are the electric and magnetic static scalar potentials Green's functions of the cavity in Coulomb gauge, respectively; and $\mathbf{\vec{G}^{\rm A}}
(\vec{r},\vec{r}^{\, \prime})$ and $\mathbf{\vec{G}^{\rm F}} (\vec{r},\vec{r}^{\, \prime})$ are the electric and magnetic dyadic potentials Green's functions of the cavity in Coulomb gauge, respectively.  The derivation of (\ref{bi-rme3d_E_general}) and (\ref{bi-rme3d_H_general}), as well as the closed form and relationships of both scalar and dyadic electric and magnetic potential Green's functions in Coulomb gauge, are cumbersome and can be found in the references cited for the BI-RME 3D theory. 

From a physical point of view, we want to remark that the surface magnetic currents $\vec{M}$ are a mathematical artifact to represent the physical connection of the waveguide-ports with the inner of the cavity (otherwise the electric tangential component on the perfect conducting walls vanishes). On the other hand, they could also be used to describe electromagnetic field discontinuities existing on surfaces separating different regions as reported in \cite{Domcke:2023qle}. In this study, we assume that the magnetic field is homogeneous in a significant region containing the cavity, and ignore the surface terms associated with the jump to a region where it vanishes. Once a set-up with a concrete magnetic field configuration is decided, our method allows us to easily include these currents in the analysis\footnote{We thank Camilo Garc\'ia-Cely and Valerie Domcke for clarifying discussions on this point.}.
%but they are not related to magnetic monopoles.

\begin{figure}[htbp]
\centering
\includegraphics[width=.9\textwidth]{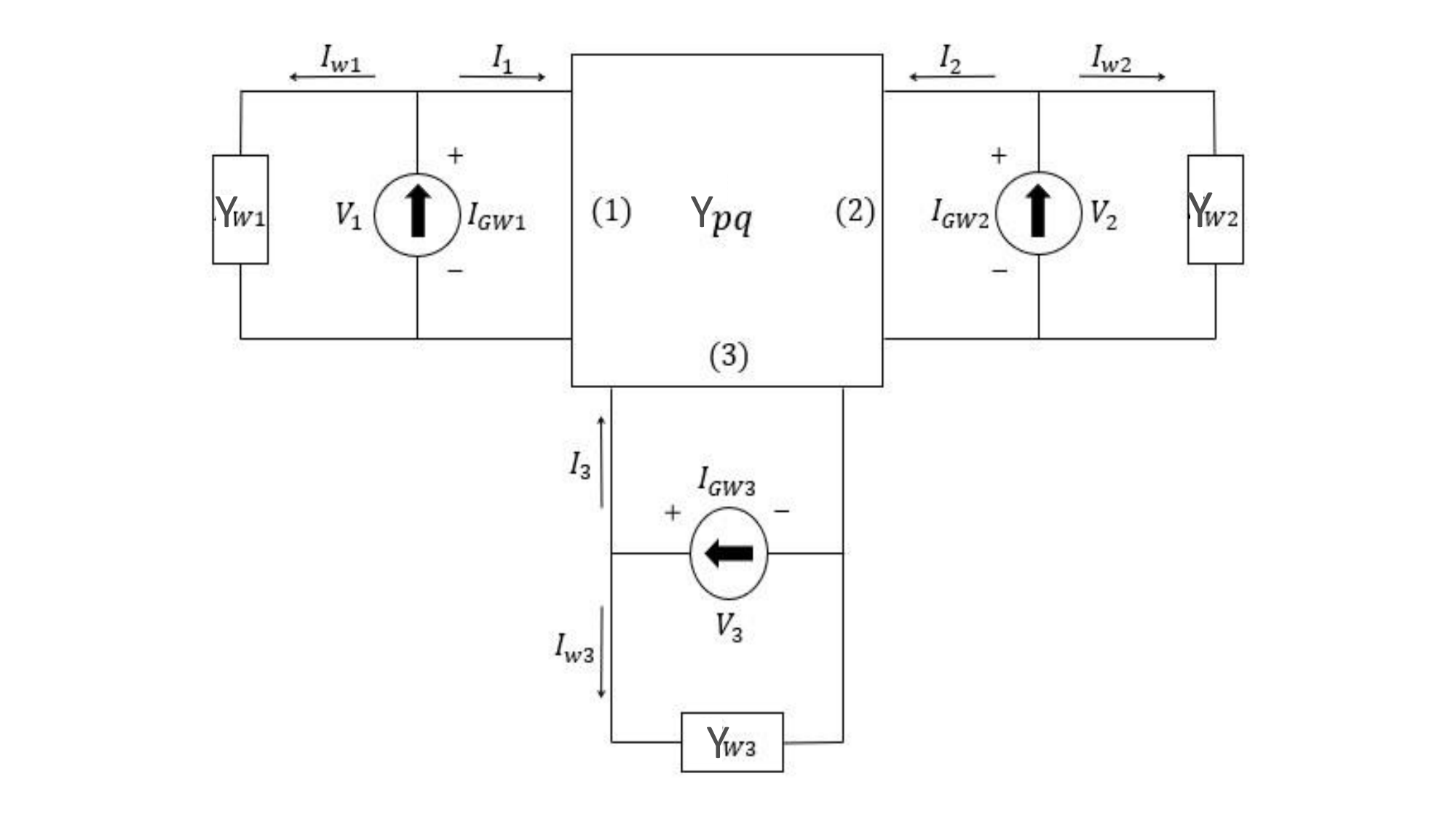}
\qquad
\caption{Scheme of a cavity with $P = 3$ ports driven by the GWs current sources $I_{GW_i}$. These currents are divided into two types of currents: the currents that drive the resonator $I_i$, and the currents injected to the ports $I_{W_i}$ which allow for the extraction of electromagnetic energy from the cavity and, as a consequence, the detection of the GW. $Y_{W_i}$ represents the wave admittance of the waveguide-port $(i)$. The index $i$ has values $i \in \{1, 2, 3\}$.}
\label{scheme_3ports}
\end{figure}
At this point, we will use the parallelism existing between the detection of dark matter axions and GWs in the context of the BI-RME 3D theory. For such purpose, we will use the theory developed in \cite{bi-rme3d_axions} for the rigorous electromagnetic analysis of the dark matter axion-photon coupling existing in a microwave cavity under the presence of an intense static magnetic field $\vec{B}_0$. In that work, some of us detailed the successful application of the BI-RME 3D method to a time-harmonic equivalent axion electric current density $\vec{J}$, which was treated as an external source. For the GWs, the formulation is similar so we will omit the mathematical derivation. Finally, the BI-RME 3D method states that the transformation of the GWs into electromagnetic energy can be formulated in terms of a set of $P$ time-harmonic current sources $I_{{GW}_i}$ exciting the cavity, which is represented by its admittance matrix $Y_{p,q}$  \cite{collin_FMI,pozar} as can be seen in figure~\ref{scheme_3ports} for the case of $P=3$ ports. In this first approach, we have neglected the effect of the higher-order modes excited in the waveguide ports, considering only the fundamental mode. The expression of the current sources is given by the following formula,
\begin{eqnarray}     
\label{I_GW_general}
	I_{GW_{i}} \, = \, \sum_{m=1}^{M}  \, \frac{\kappa_m}{k^2 - \kappa_m^2} \, \underbrace{\left( \int_{S(i)} \vec{H}_m(\vec{r})  \cdot \vec{h}_1^{(i)}(\vec{r}) \, dS \right)}_{COUPLING: CAV-PORT} \,   \,  \underbrace{\left( \int_ V \vec{E}_m(\vec{r}^{\, \prime}) \cdot \vec{J}_{GW}(\vec{r}^{\, \prime})  \, dV'   \right)}_{COUPLING: GW-CAV} \,, 
\end{eqnarray}
for $ i \in \{1, 2, 3,... P\}$, and   
where $m$ refers to the resonant modes of the cavity, $M$ being the total number of the set of resonant modes considered in the summation; $\kappa_m$ is the perturbed wavenumber due to the finite electric conductivity $\sigma$ of the metallic walls \cite{vanbladel,collin_FMI,collin_FTGW} given by
\begin{eqnarray}
\kappa_m  \, \approx  \,   k_m \, \left(1 - \frac{1}{2 \, Q_m}\right) \, + \, \mathrm{i} \, \frac{k_m}{2 \, Q_m}\,,
	 \label{eq:losses}
\end{eqnarray}
$Q_m$ being the unloaded quality factor of the $m$-th resonant mode, and $k_m$ its unperturbed wavenumber. Finally, in expression (\ref{I_GW_general}), $\Vec{E}_m$ and $\Vec{H}_m$ are the normalized electric and magnetic solenoidal eigenvectors of the cavity corresponding to the eigenfactor $k_m$, as reported in \cite{bi-rme3d_axions}; and $\vec{h}_1^{(i)}$ represents the magnetic field of the first mode (fundamental mode) of the waveguide-port $(i)$, which has to be properly normalized as described in \cite{bi-rme3d_axions}. Now, it is important to remark that the first integral of (\ref{I_GW_general}) is a surface integral performed over the access waveguide-port surface $S(i)$, which accounts for the coupling between the cavity and the port $(i)$.  The second integral is a volume integral performed over the entire volume of the cavity $V$, which is directly related to the coupling between the $m$-th resonant mode and the current $\vec{J} = \vec{J}_{GW}$ induced by the GW.

The BI-RME 3D formulation allows us to analyze the excitation of a microwave cavity by a GW through a full-wave modal technique, as it is shown in figure~\ref{scheme_3ports}, where the current sources $I_{GW_{1}}$, $I_{GW_{2}}$ and $I_{GW_{3}}$ inject electromagnetic energy generated by the GW to both the cavity (represented by its admittance matrix) as well as to the external three ports. The energy injected into the cavity will be dissipated by the Joule effect (Ohmic losses), whereas the energy delivered to the ports might allow for the detection of the GW. We will also demonstrate that information about the amplitude and the phase of the detected signals in all the ports can be calculated with the present technique, contrary to conventional methods based on a figure of merit, which only provides information about the power of the signal generated by the GW.

To proceed, we will apply the Kirchhoff laws for each waveguide-port of  figure~\ref{scheme_3ports}, resulting in
\begin{eqnarray}
\label{kirchhoff}
I_{GW_i} \, = \, I_i \, + I_{W_i} \, ;  \, V_i \, = \,  Z_{W_i} \, I_{W_i}  \, \Longrightarrow \, I_i \, = \, I_{GW_i} \, - \, I_{W_i} \, = \, I_{GW_i} \, - \, Y_{W_i} \, V_i, 
\end{eqnarray}
for $i \in \{1, 2, 3\}$, 
where $Z_{W_i}$ and $Y_{W_i}$ are the wave impedance and admittance of the waveguide-port $(i)$, respectively, which are related by $Y_{W_i} = 1/Z_{W_i}$. The classical Microwave Network Theory \cite{collin_FMI,pozar} relates the voltages at the waveguide ports $V_i$ and the currents entering into the cavity $I_i$ of a microwave linear passive component (as a cavity) with the multi-port single-mode admittance matrix $Y_{pq}$ that for the case of $P=3$ ports is given by
\begin{eqnarray}
\label{admittance_matrix}
\begin{pmatrix} I_1 \\ I_2  \\ I_3 \end{pmatrix} \, = \, 
\begin{pmatrix}
Y_{11} & Y_{12} & Y_{13}\\
Y_{21} & Y_{22} & Y_{23}\\
Y_{31} & Y_{32} & Y_{33}
\end{pmatrix} 
\, \cdot \,
\begin{pmatrix} V_1 \\ V_2  \\ V_3\end{pmatrix} .
\end{eqnarray}
By inserting (\ref{kirchhoff}) into (\ref{admittance_matrix}) and after simple mathematical manipulations, we can obtain the relationship between the unknown voltages $V_i$ and the GW current sources $I_{GW_i}$,
\begin{eqnarray}
\label{admittance_matrix_charged}
\begin{pmatrix} I_{GW_1} \\ I_{GW_2}  \\ I_{GW_3}  \end{pmatrix} \, = \, 
\begin{pmatrix}
Y_{11} + Y_{W_1} & Y_{12} & Y_{13}\\
Y_{21} & Y_{22} + Y_{W_2} & Y_{23}\\
Y_{31} & Y_{32} & Y_{33} + Y_{W_3}
\end{pmatrix}
\, \cdot \,
\begin{pmatrix} V_1 \\ V_2  \\ V_3 \end{pmatrix} \, ,
\end{eqnarray}
which is a linear system that allows one to obtain the modal voltages $V_i$ at the cavity ports. 

Finally, we can calculate the extracted power $P_{W_i}$ in each port as, 
\begin{eqnarray}
\label{power_general_waveguide}
P_{W_i}  \, = \,  \frac{1}{2} \, \mathrm{Re}(V_i \, I_{W_i}^*) \, = \, \frac{1}{2} \,\mathrm{Re}(Y_{W_i}^*) \, |V_i|^2\,,
\end{eqnarray}
where $^*$ denotes complex conjugate.

%%%%%%%%%%%%%%%%%%%%%%%%%%%%%%%%%%%%%%%%%%%%%%
\section{Application: design of a cubic cavity for GWs detection}
\label{sec:application}
%%%%%%%%%%%%%%%%%%%%%%%%%%%%%%%%%%%%%%%%%%%%%%

\begin{figure}[htbp]
\centering
\includegraphics[width=.6\textwidth]{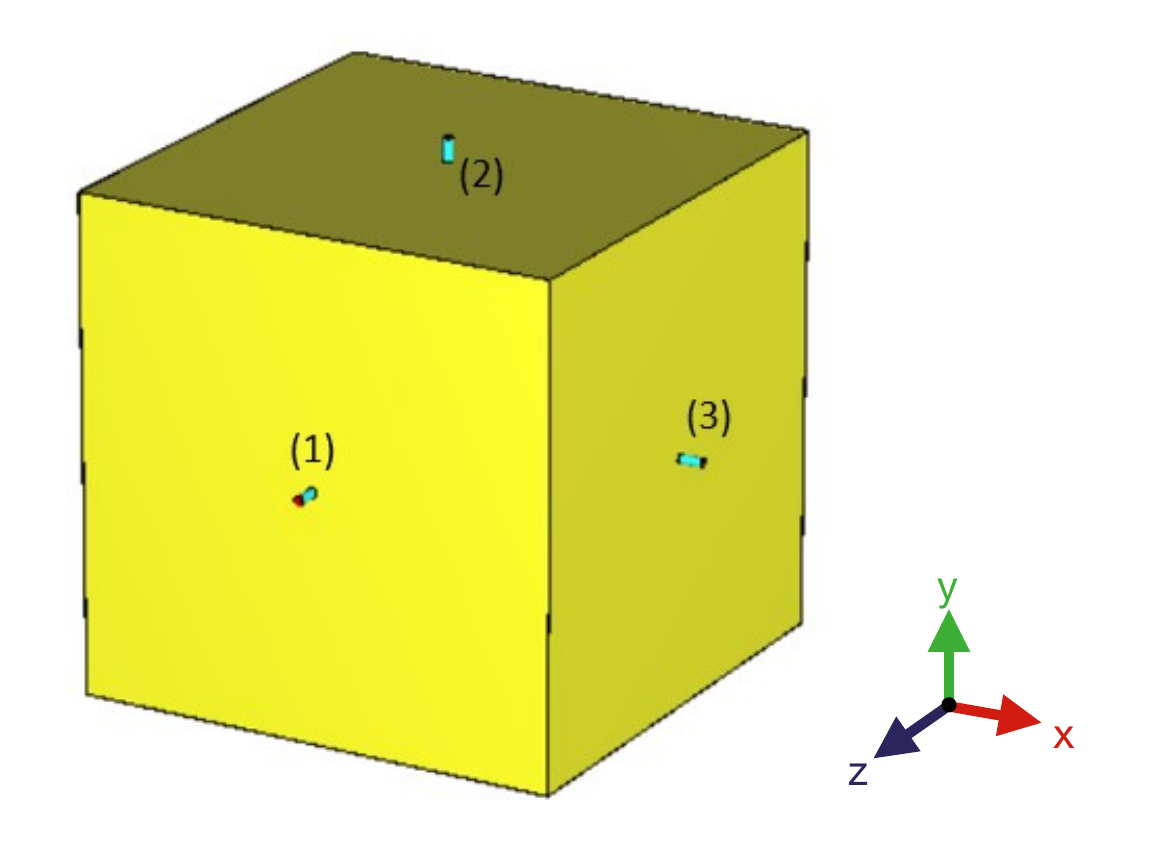}
\qquad
\caption{The cubic resonator has three perpendicular coaxial probes for the simultaneous detection of the three degenerate modes $TE_{011}$, $TE_{101}$ and $TM_{110}$ oriented along the $X$, $Y$ and $Z$ axes, respectively. These modes will be detected with the coaxial waveguide ports $(3)$, $(2)$ and $(1)$, respectively. The coaxial antennas are centered on their respective sides of the cube. The origin of the Cartesian reference system is placed in the geometrical center of the cavity; the external magnetostatic field is oriented along the $z$ axis.}
\label{scheme_cubic_cavity}
\end{figure}

During the last years, several researchers have studied the possibility of using microwave haloscopes designed for dark matter axions detection to search for GWs, making use of already existing experimental facilities, see e.g. \cite{Herman:2020wao,Berlin:2021txa,Domcke:2023bat}. In this work, we have decided to go one step further by proposing the {design of a novel cavity for the detection of GWs}. Since the coupling of a GW with a resonator is quite cumbersome, we propose to use a {cubic cavity} because it allows for the simultaneous detection of three degenerate resonant modes using three mutually perpendicular coaxial antennas, as it has been depicted in figure~\ref{scheme_cubic_cavity}.
The use of a cubic cavity may not seem ideal because of the possible 
losses of the corners, though this can be overcome by smoothing them using rounded corners \cite{rounded_corners}. Also, the coupling of a GW to a cubic geometry may not be the most efficient one. Despite these aspects, the simplicity of the design and simulations together with the possibility of using degenerate modes overcome these caveats and justify our choice of a cubic cavity. We leave a full optimization of the cavity design to future work. The selected set of degenerate modes are the $TE_{101}$, $TE_{011}$ and $TM_{110}$, whose resonant frequencies are
\begin{eqnarray}
f_{TE_{101}}  \, = \,  f_{TE_{011}} \, = \, f_{TM_{110}} \, = \, \frac{c}{\sqrt{2} \, a}\,,
\nonumber     
\end{eqnarray}
where $a$ is the edge length of the cube. It is evident that, in the absence of a GW, the three degenerate modes are identical from a physical point of view due to the symmetry of the cubic resonator.

As discussed in section~\ref{sec:gw-em}, the GWs generate a certain electromagnetic current in the presence of an external intense magnetostatic field $\vec{B}_0$.  From  \eqref{eq:max_mod}, the associated volumetric electric current density can be written as
\begin{eqnarray}   \label{j_plus_cross} 
	\vec{J} =\vec{J}_{GW}(\vec{r}) \, = \, \frac{1}{\mu_0} \, B_0 \, \left(h_+ \, \vec{J}_+(\vec{r}) \, + \, h_{\times} \, \vec{J}_{\times}(\vec{r})\right),
\end{eqnarray}	
where the two components are the currents associated with the cross ($\times$) and plus ($+$) polarizations present in \eqref{eq:TT}.
Once written in SI units, this suffices to describe the presence of GWs within a microwave resonator in the BI-RME 3D scenario. Here and in the following we assume that the external magnetostatic field is homogeneous and oriented in the $Z$ axis of a Cartesian reference system centered in the center of mass of the cavity, $\vec{B}_0 \, = \, B_0 \, \hat{z}$.
%$\vec{J}_+$ and $\vec{J}_{\times}$ describe the spatial profile and polarization of the GW. 

%
\begin{figure}[htbp]
\centering
\includegraphics[width=.9\textwidth]{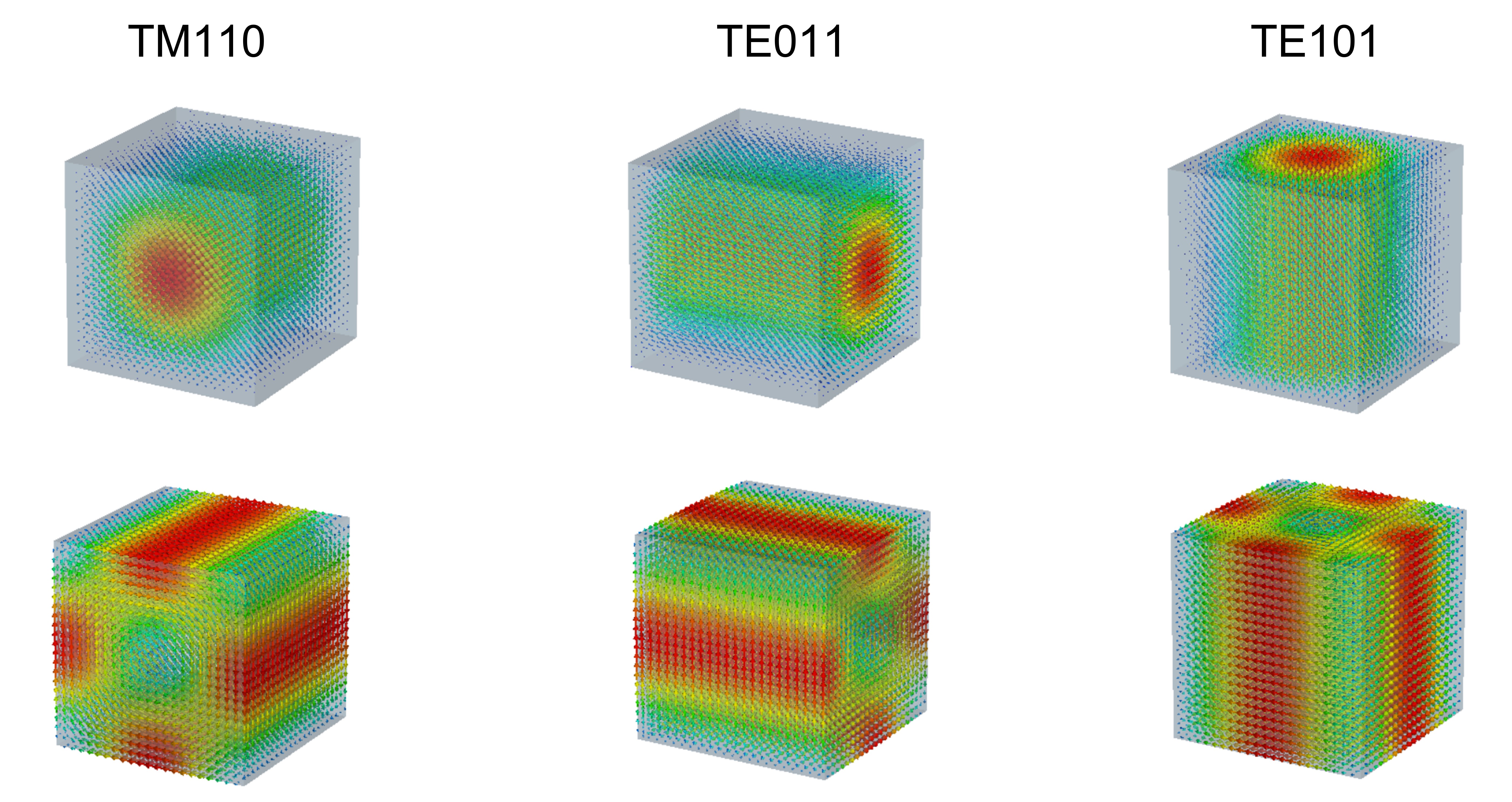}
\qquad
\caption{Electric (up) and magnetic (down) field distributions for the three considered modes $TM_{110}$, $TE_{011}$ and $TE_{101}$ of   cavity C1. The electromagnetic field distributions are identical in the other cavities C2 and C3. The red hue represents the largest values for the modules of electric and magnetic fields, while the blue color represents the lowest values for these modules. The remaining colors in this illustration represent intermediate levels between minima and maxima.}
\label{electric_magnetic_fields}
\end{figure} 

In this section, we will apply the BI-RME 3D technique described in section \ref{sec:bi-rme3d_formulation} to the accurate and efficient characterization of a cubic cavity. To explore different ranges of frequencies, we have studied three cavities tuned at $100$ MHz (Cavity 1: C1), $1$ GHz (Cavity 2: C2), and $10$ GHz (Cavity 3: C3). Numerical simulations in this section have been made with the commercial software CST Studio \cite{CST}, and post-processed with Matlab \cite{MATLAB}. Table~\ref{tab:cavities} summarizes the characteristics of the cavities. The electromagnetic field distributions of these modes have been plotted in figure~\ref{electric_magnetic_fields}. The unloaded quality factors $Q_m$ of the three modes have been calculated \cite{collin_FMI} with the electric conductivity of copper at cryogenic temperature, $\sigma = 2 \cdot 10^9$ S/m. From previous experience in the design and manufacturing of high-$Q$ resonant cavities, it is expected a moderate reduction ($\thicksim30\%$) of the unloaded quality factor in the actual cavity as regards simulations, mainly because of losses associated with walls roughness at very low temperatures and possible leakage in slits between walls. To attenuate this issue, roughness can be greatly reduced by electro-polishing, and walls can be appropriately soldered. Moreover, as we comment in the conclusions section, there is room for improvement in the quality factor using magnetic-resilient superconducting materials. The coaxial connectors used for the line-cavity coupling probes are BNC for C1, and SMA for C2 and C3; their inner radii $r_i$, outer radii $r_o$, relative dielectric permittivity $\varepsilon_r$ and the penetration distance of the coaxial antenna inside the cavity $d$ for critical coupling regime are also reported in table \ref{tab:cavities}. The reflection scattering parameters at the resonance frequencies are around $|S_{11}| = |S_{22}| = |S_{33}| \approx -40$ dB for the three antennas of the three cavities, which indicates a good level of critical coupling regime. The characteristic coaxial impedance $Z_0$ of the three coaxial probes of each cavity is given by
\begin{eqnarray}
Z_0  \, = \, \frac{1}{Y_0} \, = \,  Z_W \, \frac{\ln{(r_o/r_i)}  }{2 \, \pi} \, \, ; \; \; Z_W \, = \, \frac{1}{Y_W} \, = \, \sqrt{\frac{\mu_0}{\varepsilon_0 \, \varepsilon_r}} \, ,
\nonumber     
\end{eqnarray}
where $Y_0$ is the characteristic coaxial admittance, and $Z_W$ and $Y_W$ are the wave coaxial impedance and admittance, respectively. The impedance of both BNC and SMA coaxial connectors is $Z_0 \, = \, 50 \, \Omega$.

From an experimental point of view, it is important to remark that the complex voltage (phasor) measured in the coaxial waveguide ports $v_{i}$ can be easily calculated as a function of the waveguide ports voltages $V_i$ resulting in \cite{bi-rme3d_axions},

\begin{eqnarray}  
\label{voltage_measured}
v_i \, = \, \sqrt{\frac{\ln{(r_o/r_i)}}{2 \, \pi}} \, V_i\,,
\end{eqnarray}
which allows one to rewrite (\ref{power_general_waveguide}) as
\begin{eqnarray}
\label{power_coaxial}
P_{W_i}  \, = \, \frac{1}{2} \, Y_{0} \, |v_i|^2 \ .
\end{eqnarray}

{A final remark in the cavity design and operation is the need for a moving mechanism at each probe (monopole antenna) to allow its introduction or recession in the cavity. This is necessary because, due to manufacturing tolerances, the resonant frequencies of the three degenerated modes will not be exactly equal. With this movement the corresponding mode is slightly perturbed and, therefore, its resonant frequency can reach the desired value. Moreover, movable probes allow to de-degenerate the three modes, if needed, and to modify the coupling between the coaxial line and the cavity.}

\begin{table}[htbp]
\centering
\begin{tabular}{|c|c|c|c|}
\hline
& CAVITY 1 & CAVITY 2 & CAVITY 3\\
\hline
$a$ (mm) & 2119.85 & 211.98 & 21.19\\
\hline
$Q_{TE101}$ & $6.27 \cdot10^5$ & $1.98 \cdot 10^5$ & $6.25 \cdot 10^4$ \\
\hline
$Q_{TE011}$ & $6.27 \cdot 10^5$ & $1.98 \cdot 10^5$ & $6.25 \cdot 10^4$\\
\hline
$Q_{TM110}$ & $6.27 \cdot 10^5$ & $1.98 \cdot 10^5$ & $6.25 \cdot 10^4$\\
\hline
$r_i$ (mm) & 7.00 & 0.0635 & 0.0635\\
\hline
$r_o$ (mm) & 16.00 & 0.211 & 0.211\\
\hline
$\varepsilon_r$ & 1.00 & 2.08 & 2.08\\
\hline
$d$ (mm) & 32.80 & 5.30 & 0.21\\
\hline
\end{tabular}
\caption{Characteristics of the three cubic cavities and their coaxial probes. \label{tab:cavities}}
\end{table}

\subsection{Coupling form factor}

\begin{figure}[htbp]
\centering
\includegraphics[width=0.8\textwidth]{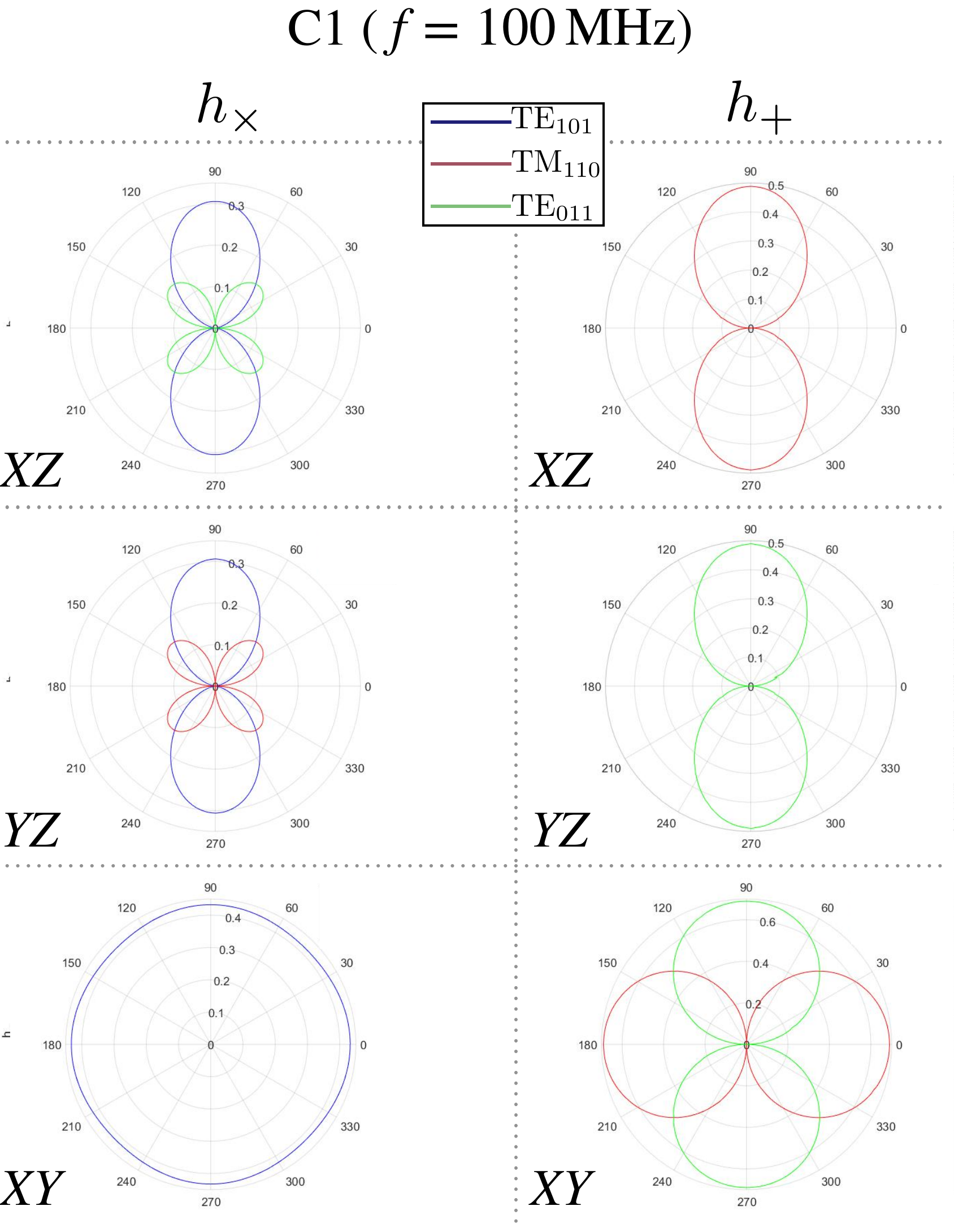}
\caption{Form factor $\tilde\eta_{m_{+,\times}}$ between the GW and the three degenerate resonant modes as a function of the GW incidence angular direction for the cavity C1 ($f \, = \, 100$ MHz). The polar angle is expressed in degrees. Some curves cannot be seen because the form factor is negligible compared to the rest of the results. Left: cross-polarization; Right: plus polarization. Up: GW incidence in the $XZ$ plane; Center: GW incidence in the $YZ$ plane; Down: GW incidence in the $XY$ plane.}
\label{coupling_XZ_YZ_XY_C1}
\end{figure}

\begin{figure}[htbp]
\centering
\includegraphics[width=0.8\textwidth]{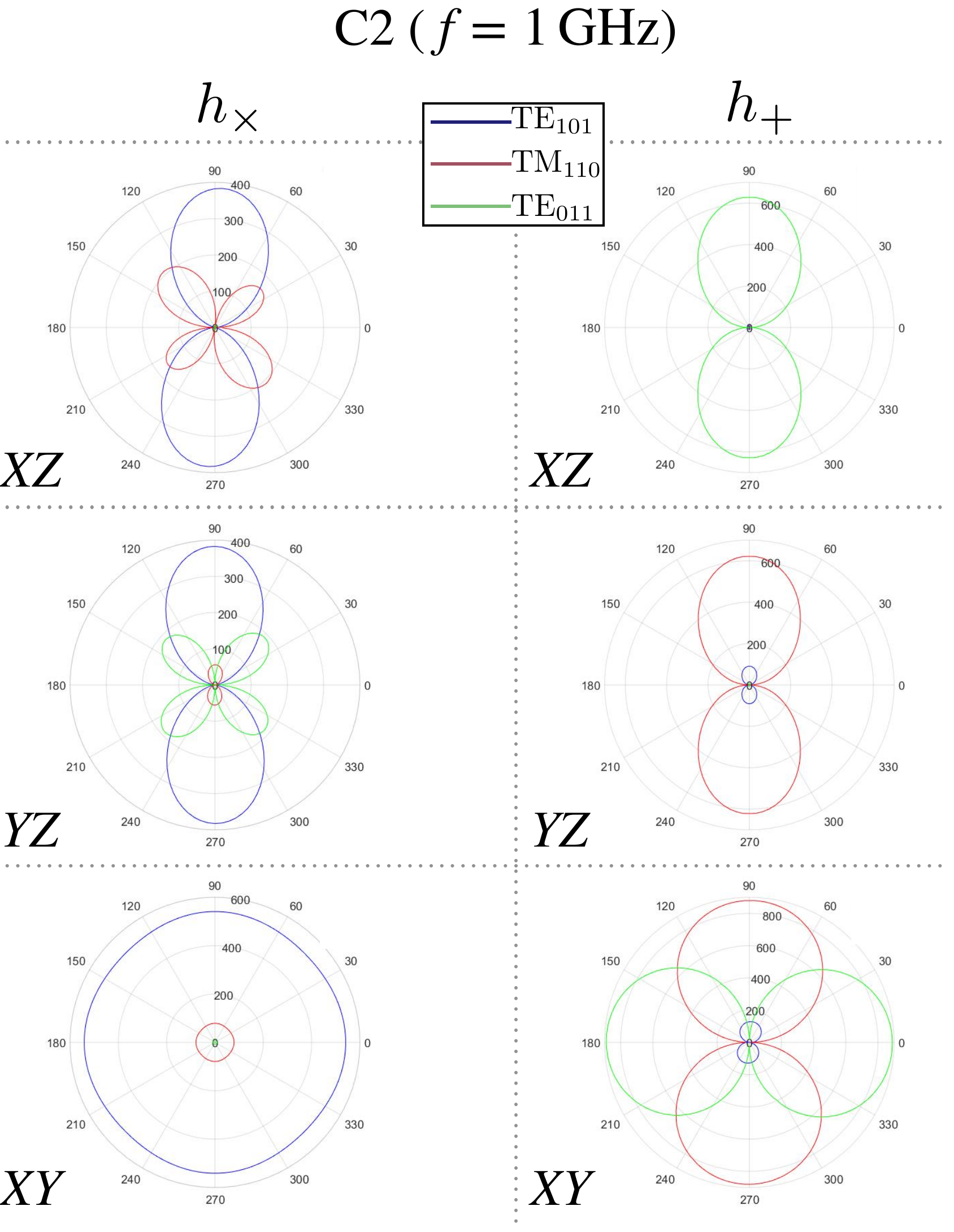}
\caption{Form factor $\tilde\eta_{m_{+,\times}}$ between the GW and the three degenerate resonant modes as a function of the GW incidence angular direction for the cavity C2 ($f \, = \, 1$ GHz). The polar angle is expressed in degrees.  Some curves cannot be seen because the form factor is negligible compared to the rest of the results. Left: cross-polarization; Right: plus polarization. Up: GW incidence in the $XZ$ plane; Center: GW incidence in the $YZ$ plane; Down: GW incidence in the $XY$ plane.}
\label{coupling_XZ_YZ_XY_C2}
\end{figure}

\begin{figure}[htbp]
\includegraphics[width=0.8\textwidth]{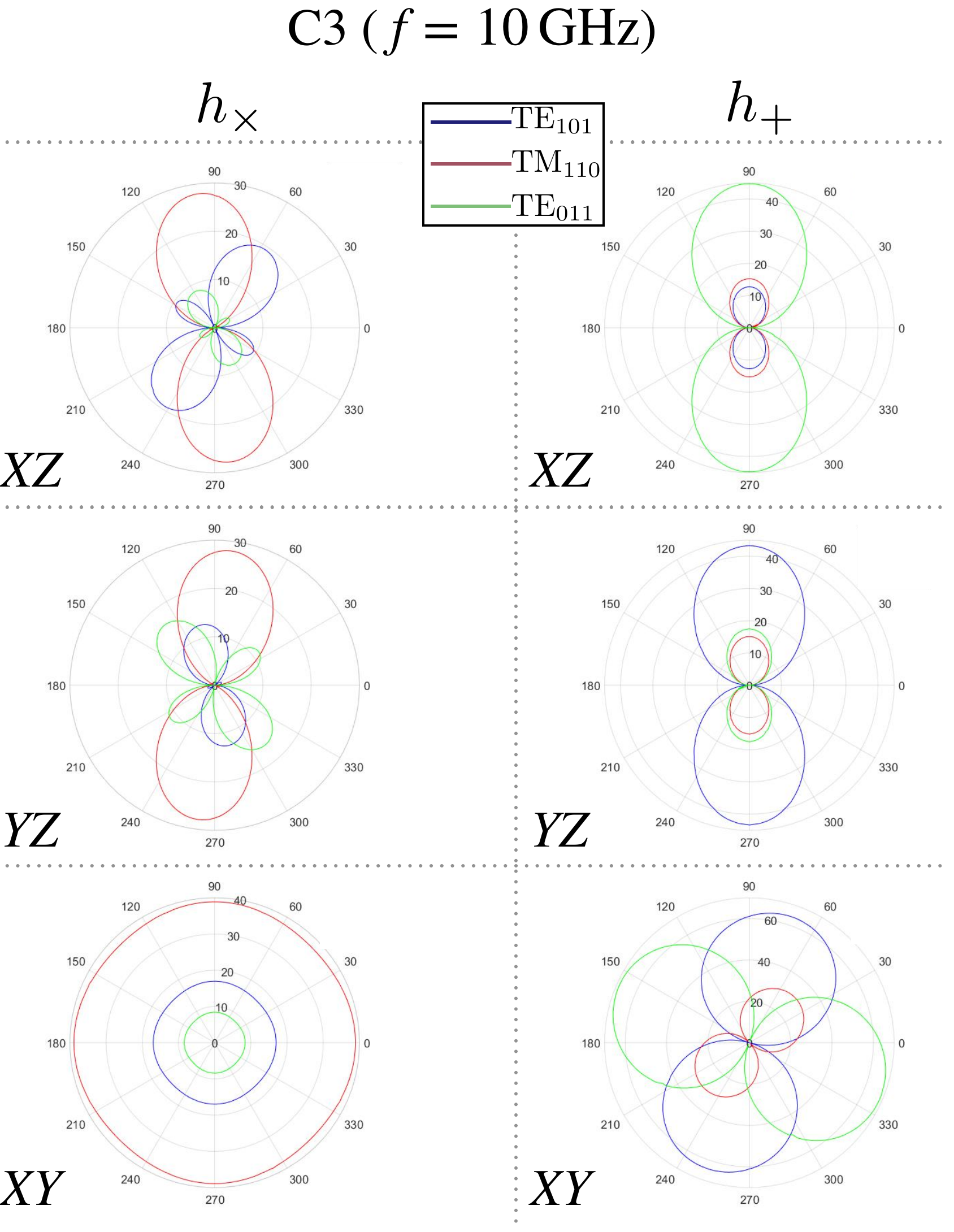}
\caption{Form factor $\tilde\eta_{m_{+,\times}}$ between the GW and the three degenerate resonant modes as a function of the GW incidence angular direction for the cavity C3 ($f \, = \, 10$ GHz). The polar angle is expressed in degrees.  Some curves cannot be seen because the form factor is negligible compared to the rest of the results. Left: cross polarization; Right: plus polarization. Up: GW incidence in the $XZ$ plane; Center: GW incidence in the $YZ$ plane; Down: GW incidence in the $XY$ plane.}
\label{coupling_XZ_YZ_XY_C3}
\end{figure}

To analyze the coupling between the GWs and the resonant modes, we have used the definition of the dimensionless form factor between the GW and the $m$ mode introduced in \cite{Berlin:2021txa},
\begin{eqnarray}  \label{coupling_coefficient_GW_Em}
\tilde\eta_{m_{+,\times}} \, = \, \frac{\left| \int_V \Vec{E}_m(\vec{r}) \, \cdot \, \vec{J}_{+,\times}(\vec{r}) \, dV \right|}{V^{1/2} \, \left| \int_V \Vec{E}_m(\vec{r}) \, \cdot \, \Vec{E}_m(\vec{r}) \, dV \right|^{1/2}} \,,
\end{eqnarray}
where the integral of the denominator is equal to one because of the orthonormalization condition used in BI-RME 3D \cite{conciauro}; $V$ is the volume of the cavity. This parameter has been represented in polar coordinates as a function of the incidence angle $\theta$ of the GW with respect to the $Z$ axis for waves
propagating along the $XZ$, $YZ$ planes ($\theta = 0^{\circ}$ refers to the $Z$ axis) and the angle with respect to the $X$ axis 
for waves propagating in the $XY$ plane ($\theta = 0^{\circ}$ direction refers to the $X$ axis in this case). The value of $\tilde\eta_{m_{+,\times}}$ is shown in terms of $\theta$ in figures~\ref{coupling_XZ_YZ_XY_C1},~\ref{coupling_XZ_YZ_XY_C2} and~\ref{coupling_XZ_YZ_XY_C3} for the C1, C2 and C3 cavities, respectively. As clear from these plots, the coupling strongly depends on the direction of the incidence plane, the polarization, and the operation frequency.

It is important to remark here that this form factor is of a very different nature than the form factor in dark matter axion detection. Whilst the latter does not depend on any axion intrinsic parameter, in the case of the graviton, the form factor depends on the complex relationship of the induced current ($\vec{J}_{GW}$) with frequency. This makes this factor not normalized and implies that changes with frequency are not just due to the cavity geometry, but also to the frequency dependence of the GW-induced current. This explains that figures ~\ref{coupling_XZ_YZ_XY_C1} -~\ref{coupling_XZ_YZ_XY_C3} show different magnitude scales. 

It is also worth noting that, unlike the axion haloscope, a cavity for GW detection can work with modes whose electrical field is not aligned with the external magnetic field. This can be observed, for instance, in figure~\ref{coupling_XZ_YZ_XY_C1} for the cross-polarization case, where the $TE_{101}$ ($\vec{E} \, = \, E_y \, \hat{y}$) gets the best form factor in most angles of incidence. This allows for more freedom in selecting resonant modes and opens up novel opportunities for the design of microwave-cavity gravitational-wave detectors.

Moreover, figures ~\ref{coupling_XZ_YZ_XY_C1} -~\ref{coupling_XZ_YZ_XY_C3} also provide two crucial findings. The first one is that gravitational wave energy is transformed into currents inside the cubic cavity, which excite the three orthogonal degenerate modes. Thus, utilizing a probe for each of these degenerate modes and the right mix of the extracted signals allows for improved detection sensitivity. Because the gravitational wave stimulates all degenerate modes at the resonant frequency, omitting to combine signals would result in a loss of sensitivity. Alternatively, we can profit from this multiple coupling of the GW by slightly detuning the three degenerate modes (for instance by proper movement of small mechanical parts in the cavity). In this way, the three modes are no longer degenerate, allowing the GW detection in three different and very close frequencies. The same applies to higher-order modes, provided that their form factors are adequate for detection.

A further finding is that combining the degenerate modes in the proposed cubic cavity results in some cases in increasing the range of incident angles that can be detected (see figures~\ref{coupling_XZ_YZ_XY_C1}~-~\ref{coupling_XZ_YZ_XY_C3}, plane $XY$, plus polarization case). From a practical point of view, the preferred mode for detection will be that with larger values of the form factor and which shows a more isotropic behavior to be able to detect GWs impinging from different directions. Table~\ref{tab:optimal_GW_C1} shows the proper modes following this criterion for cavity C1. The best behavior is found for GWs in the $XY$ plane for the three cavities. In this case, only one mode is necessary for covering all possible angles for cross-polarization.  Nevertheless, the plus polarization requires a combination (sum) of two modes to avoid blind angles. In the other planes, the combination of modes does not avoid completely blind angles. That is the case for  $\theta={0, \pi}$ in the $XZ$ and $YZ$ planes for both polarizations. In these cases there are different options to obtain a proper form factor for these angles, such as rotating the cavity inside the magnet to modify the magnetostatic field direction with regard to the cavity coordinate system, using a second rotated cavity, or, for long enough signals, one can profit from the rotation of the Earth to achieve this change.

\begin{table}[htbp]
\centering
\begin{tabular}{|c|c|c|c|}
\hline
     & XZ plane & YZ plane & XY plane  \\
\hline
$\times$ polarization & $TE_{101}$ & $TE_{101}$ & $TE_{101}$ \\
\hline
$+$ polarization & $TM_{110}$ & $TE_{011}$ & $TE_{011}$/$TM_{110}$ \\
\hline
\end{tabular}
\caption{Optimal mode for GW detection depending on polarization and plane of incidence for cavity C1.}
\label{tab:optimal_GW_C1}
\end{table}

To compute the detected power at each port, we have assumed that the GW impinges in the angle $\theta_0$ which generates the maximum coupling between the GW and the cavity mode, as reported in table~\ref{tab:GW_angles}.
\begin{table}[htbp]
\centering
\begin{tabular}{|c|c|c|c|}
\hline
& CAVITY 1 (C1) & CAVITY 2 (C2) & CAVITY 3 (C3)\\
\hline
$XZ$ plane & $90.0 \, \, \, (+)$ & $90.0 \, \, \, (+)$ & $90.0 \, \, \, (+)$ \\
\hline
$YZ$ plane & $90.0 \, \, \, (+)$ & $90.0 \, \, \, (+)$ & $90 .0\, \, \, (+)$ \\
\hline
$XY$ plane & $0.0 \, \, \, (+)$ & $0.0 \, \, \, (+)$ & $70.2 \, \, \, (+)$ \\
\hline
\end{tabular}
\caption{Incidence angle $\theta_0$ (degrees) for maximum coupling between the GW and the resonant modes. The GW polarization for the maximum coupling is indicated.} \label{tab:GW_angles}
\end{table}
\begin{table}[htbp]
\centering
\begin{tabular}{|c|c|c|c|c|c|c|c|}
\hline
 Cavity  & V (L) & Magnet & $B_0$ (T) & $T_{phys}$ (mK) & $T_{sys}$ (K) & $\Delta f$ (kHz) \\
\hline
C1 & 9526.1056 & KLASH \cite{KLASH} & 0.6 & 4500 & 8 & 5 \\
\hline
C2  & 9.5243 & CAPP \cite{magnet_C2_C3} & 12 & 30 & 1 & 10 \\
\hline
C3  & 0.0095 & CAPP \cite{magnet_C2_C3} & 12 & 30 & 1 & 20 \\
\hline
\end{tabular}
\caption{Characteristics of the magnets and parameters for the data acquisition system.}
\label{tab:magnets}
\end{table}


\subsection{Sensitivity analysis}

In this first analysis of the problem, we have neglected the inter-coupling effect of the three coaxial probes of each cavity, which is a good approach given that the level of the transmission scattering parameters at the resonant frequencies is around $|S_{21}| = |S_{31}| = |S_{32}| \approx -50$ dB for the three coaxial antennas of the three cavities. As a consequence, we will neglect the mutual coupling among the three probes of each cavity, and assume that they operate independently. Thus, we do not need to solve the linear system represented in (\ref{admittance_matrix_charged}). Consequently, the power detected at each port can be expressed as
\begin{eqnarray}
\label{P_wi}
P_{W_{i_{\times,+}}}  =  \frac{1}{2} \, |h_{\times,+}|^2 \, \frac{B_0^2 \, V}{\mu_0^2} \, \frac{{\rm Re}(Y_{W_i})}{|{Y_{W_i} \, + \, Y_{ii}|}^2} \, {\left| \, \sum_{m=1}^{M}  \, \frac{\kappa_m}{k^2 - \kappa_m^2} \, \tilde\eta_{m_{+,\times}} \, {\int_{S(i)} \vec{H}_m(\vec{r})  \cdot \vec{h}_1^{(i)}(\vec{r}) \, dS}  \right|}^2 ,
\end{eqnarray}
which is proportional to the GW amplitude square $|h_{\times,+}|^2$.

The Dicke radiometer equation  \cite{pozar} provides the noise power at port $i$, 
\begin{eqnarray}
\label{P_noise}
    P_{N_i} \, = \, k_B \, T_{sys_i} \, \sqrt{\frac{\Delta f}{\Delta t}}\,,
\end{eqnarray}
where $k_B$ is the Boltzmann constant, $T_{sys_i}$ is the noise temperature of the system at port $i$, $\Delta f$ is the detection bandwidth, and $\Delta t$ is the detection time. This equation allows us to calculate the exclusion limits for the sensitivity of the experiment at port $i$ given and amplitude of the GW $|h_{i_{\times,+}}|$ for a given signal-to-noise ratio $S/N \, = \, P_{W_{i_{\times,+}}}/P_{N_i}$. By inserting (\ref{P_wi}) and (\ref{P_noise}) in the definition of $S/N$, we finally obtain, for each $i$,
\begin{eqnarray}
    \label{h_i_cross_plus}
    |h_{i_{\times,+}}| ={\left( \frac{2 \, (S/N) \, k_B \, T_{sys_i}}{{\rm Re}(Y_{W_i})} \right)}^{1/2} && {\left( \frac{\Delta f}{\Delta t} \right)}^{1/4}\nonumber\\&& \times \frac{\mu_0}{B_0 \, V^{\frac{1}{2}}} \, \frac{|Y_{W_i}+Y_{ii}|}{\left| \, \sum_{m=1}^M  \, \frac{\kappa_m}{k^2 - \kappa_m^2} \, \tilde\eta_{m_{\times,+}} \, {\int_{S(i)} \vec{H}_m(\vec{r})  \cdot \vec{h}_1^{(i)}(\vec{r}) \, dS}  \right|} .
    \hspace{2cm}
\end{eqnarray}

 \subsection{Realistic sensitivities and discussion}

In this section, we compute the sensitivities $|h_{i_{\times,+}}|$ assuming that the GW impinges with the angle $\theta_0$ which generates the maximum coupling between the GW and the cavity mode, as reported in table~\ref{tab:GW_angles}. To perform realistic numerical calculations we use the data from different magnet facilities for each cavity which have been described in table~\ref{tab:magnets}. The magnet bore in these facilities is comparable to or bigger than the corresponding cavity. Other important parameters for their election are the magnetic field magnitude and the physical temperature. These three magnets are solenoids, but dipole or quasi-dipole magnets, as BabyIAXO \cite{IAXO:2020wwp}, already proposed for dark matter axion detection \cite{Ahyoune:2023gfw}, could also be used for GW and the cavity concept described here.

\begin{figure}[htbp]
\centering
\includegraphics[width=.8\textwidth]{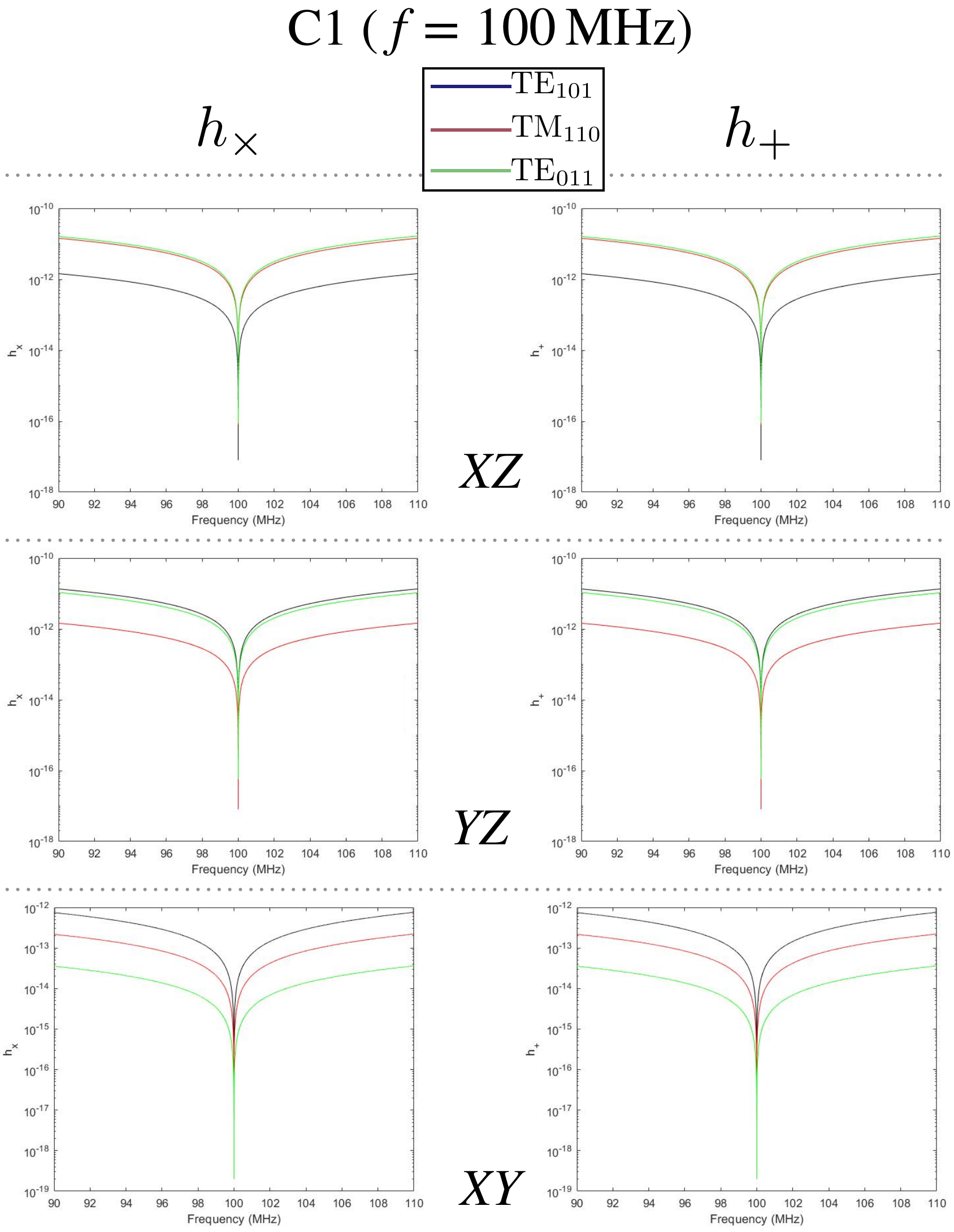}
\caption{GWs amplitudes $h_{\times}$ and $h_{+}$ as a function of frequency in the three coaxial probes of the cavity C1 ($f \, = \, 100$ MHz). Magnetostatic field: $B_0 = 0.6$ T; signal-to-noise ratio: $S/N=3$; temperature of the system: $T_{sys} = 8$ K; frequency detection bandwidth: $\Delta f = 5$ kHz;  detection time: $\Delta t = 1$ ms. Left: cross-polarization; Right: plus polarization. Up: GW incidence in the $XZ$ plane; Center: GW incidence in the $YZ$ plane; Down: GW incidence in the $XY$ plane.}
\label{h_XZ_YZ_XY_C1}
\end{figure}

\begin{figure}[htbp]
\centering
\includegraphics[width=.8\textwidth]{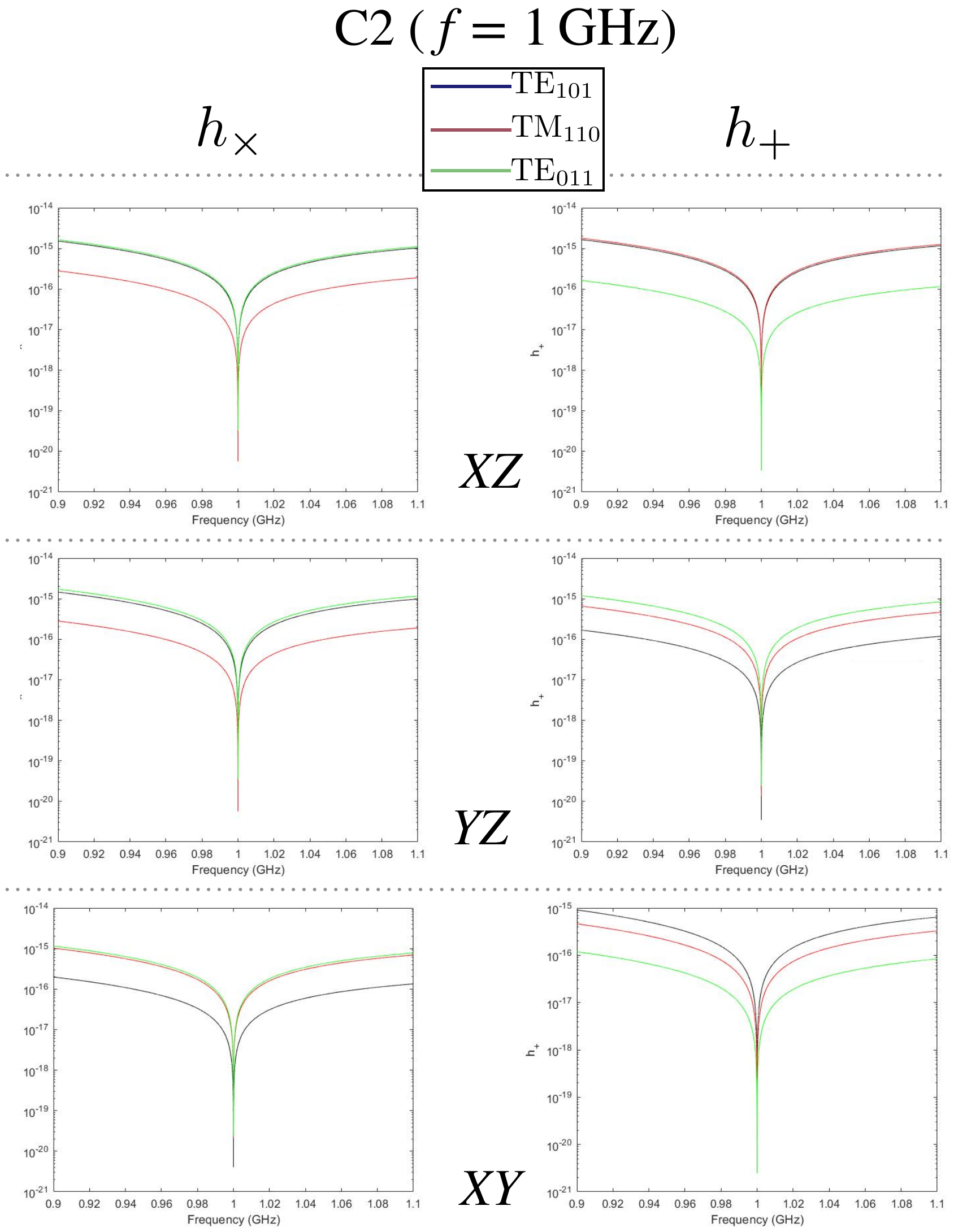}
\caption{GWs amplitudes $h_{\times}$ and $h_{+}$ as a function of frequency in the three coaxial probes of the cavity C2 ($f \, = \, 1$ GHz). Magnetostatic field: $B_0 = 12$ T; signal-to-noise ratio: $S/N=3$; temperature of the system: $T_{sys} = 1$ K; frequency detection bandwidth: $\Delta f = 10$ kHz;  detection time: $\Delta t = 1$ ms. Left: cross-polarization; Right: plus polarization. Up: GW incidence in the $XZ$ plane; Center: GW incidence in the $YZ$ plane; Down: GW incidence in the $XY$ plane.}
\label{h_XZ_YZ_XY_C2}
\end{figure}

\begin{figure}[htbp]
\centering
\includegraphics[width=.8\textwidth]{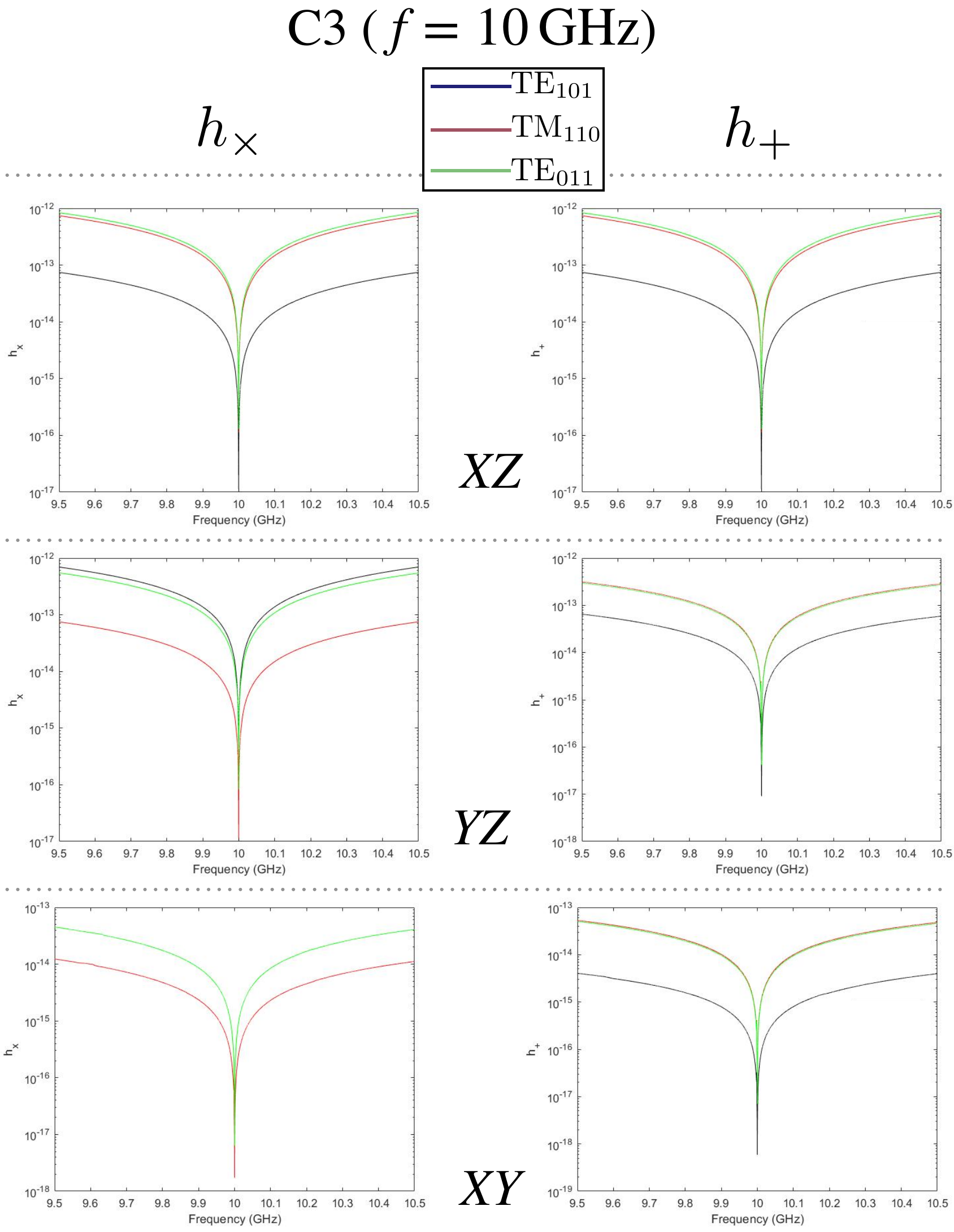}
\caption{GWs amplitudes $h_{\times}$ and $h_{+}$ as a function of frequency in the three coaxial probes of the cavity C3 ($f \, = \, 10$ GHz). Magnetostatic field: $B_0 = 12$ T; signal-to-noise ratio: $S/N=3$; temperature of the system: $T_{sys} = 1$ K; frequency detection bandwidth: $\Delta f = 20$ kHz;  detection time: $\Delta t = 1$ ms. Left: cross-polarization; Right: plus polarization. Up: GW incidence in the $XZ$ plane; Center: GW incidence in the $YZ$ plane; Down: GW incidence in the $XY$ plane.}
\label{h_XZ_YZ_XY_C3}
\end{figure}

Therefore, we will assume that the cavity C1 might be introduced in a magnet test-bed similar to KLOE (KLASH) \cite{KLASH} \cite{alesini_KLASH}, where the external magnetostatic field is $B_0 =0.6$ T and the system temperature is $T_{sys} = 8$ K. This value is calculated assuming 4 K in the cavity and an extra noise temperature added by the read-out chain of 4 K, which is mainly produced by the first amplifier, in this case a cryogenic low-noise amplifier. The detection bandwidth used in the simulations is $\Delta f = 5$~kHz, entering the cavity-loaded quality factor $Q_L = f_1/\Delta f$, $f_1 = 100$ MHz being the resonance frequency. We have used in the simulations a detection time $\Delta t = 1$~ms, and a signal-to-noise ratio $S/N = 3$ for all the cavities. This short detection time is key to accessing some of the signals that may be present in the studied frequency band \cite{Berlin:2021txa}. An important remark is that in some previous studies (as in \cite{Berlin:2021txa}) a longer integration of $1$\,s time has been considered. To compare our results with those of these studies, recall that the constraint in the maximum allowed GW amplitude grows as with $(\Delta t)^{1/4}$.
The sensitivity for the cavity C1 has been plotted in figure~\ref{h_XZ_YZ_XY_C1} using (\ref{h_i_cross_plus}), observing that the minimum detected amplitude is around $|h_{i_{\times,+}}| \approx 1 \cdot 10^{-19}$. We have also computed the sensitivity curves of the cavities C2 and C3 in figure~\ref{h_XZ_YZ_XY_C2} and figure~\ref{h_XZ_YZ_XY_C3}, respectively. The magnet chosen for these two cavities is the Oxford-Leiden one from CAPP (see table 3 for details) \cite{magnet_C2_C3}. We have assumed here that we can leverage the 30 mK of the set-up and use a Josephson parametric amplifier as the first amplifier in the read-out chain \cite{magnet_C2_C3}. Taking this into account, we expect a total noise temperature well below 1 K. For these cases the minimum detected amplitude is $|h_{i_{\times,+}|} \approx 2 \cdot 10^{-20}$ for C2, and $|h_{i_{\times,+}|} \approx 6 \cdot 10^{-19}$ for C3. It is evident that these values are far from those expected from different models of GWs in this band \cite{Domcke:2023qle,Aggarwal:2020olq}. Still, it is important to consider that those are the first steps in this emerging field, where neither the cavity design nor the readout system are optimized.  We will come back to possible improvements 
 in section~\ref{sec:conclusions}. It is also very relevant that our results are on the bulk part of other methods suggested to detect similar GWs at these high frequencies \cite{Aggarwal:2020olq,Domcke:2023qle}.

\begin{figure}[htbp]
\centering
\includegraphics[width=.9\textwidth]{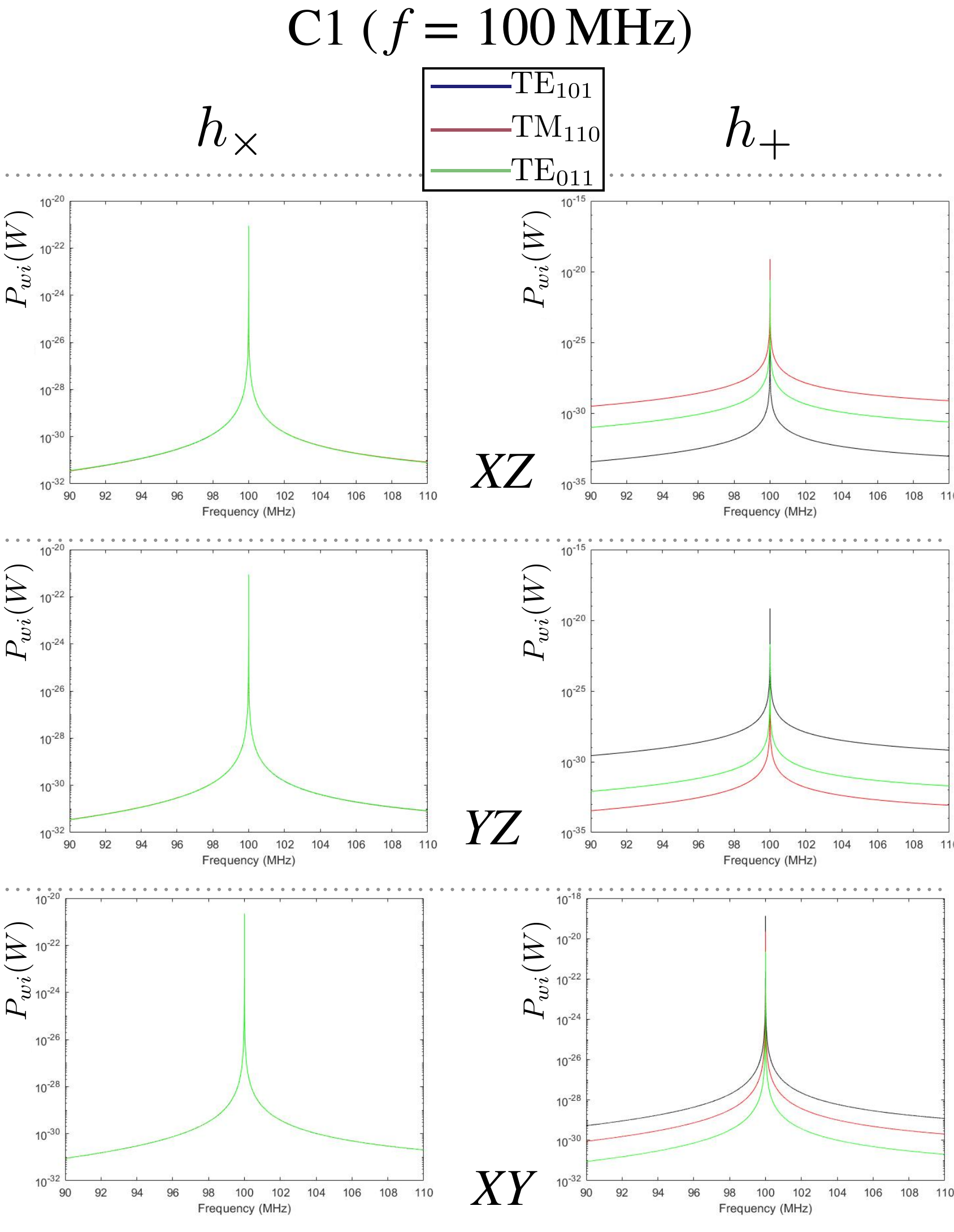}
\caption{Detected power of the GWs as a function of frequency in the three coaxial probes of the cavity C1 ($f \, = \, 100$ MHz), related to the GW amplitudes obtained in figure~\ref{h_XZ_YZ_XY_C1}. Magnetostatic field: $B_0 = 0.6$ T; signal-to-noise ratio: $S/N=3$; temperature of the system: $T_{sys} = 8$ K; frequency detection bandwidth: $\Delta f = 5$ kHz;  detection time: $\Delta t = 1$ ms. Left: cross-polarization; Right: plus polarization. Up: GW incidence in the $XZ$ plane; Center: GW incidence in the $YZ$ plane; Down: GW incidence in the $XY$ plane.}
\label{power_XZ_YZ_XY_C1}
\end{figure}

Figure~\ref{power_XZ_YZ_XY_C1} depicts the detected GWs power $P_{W_i}$ as a function of frequency in the three coaxial probes of the cavity C1 for the same optimal sensitivities in figure~\ref{h_XZ_YZ_XY_C1}. It is worth noting that, in the case of cross-polarization, the detected power level is around $10^{-21}$ W, which is consistent with the expected level of power estimated in dark matter axion studies  \cite{bi-rme3d_axions}. However, for the analyzed scenarios, the plus polarization exhibits greater observed power levels, with values around $10^{-19}$ W. This means that we are mainly sensitive to one polarization, at least with the proposed cubic cavity and incoming directions described in this work. In the future, we will study how this statement is modified in generic directions for the incoming GW.

It is important to emphasize at this point that in the frequency response computations, we have not used the classical Lorentzian approach for describing the frequency resonant curves; on the contrary, the BI-RME 3D theory provides the wide-band exact solution of the cavity electrical response. 

As indicated in section~\ref{sec:bi-rme3d_formulation}, the BI-RME 3D method is capable of computing the complex values of voltages in the cavity ports for the ranges of frequencies under study. Figures 11 and 12 show, respectively, the magnitude and phase for detected voltages, $v_{i}$, as a function of frequency in the three coaxial probes of the cavity C1 around its resonant frequency ($f = 100$~MHz). As expected from Figure 10, voltages detected from plus polarization are higher than those produced by cross-polarization. In fact, plus polarization voltages show maximum values around $10^{-6}$ V while cross polarization ones are around $10^{-7}$ V. Moreover, all of the degenerate modes and GW polarizations exhibit very similar behavior for voltage phase values, with a phase shift at the resonant frequency. For concision, the phase differences between port signals are not displayed in this paper; however, they reveal very similar behaviors in both polarizations, not showing phase shifts at the resonant frequency. Additionally, it should be noted that the detected phase values vary with GW polarization and incident planes, suggesting that GW polarizations and incident angles may be detected with this type of cavity.


%%%%%%%%%%%%%%%%%%%%%%%%%%%%%%%%%%%%%%%%%%%%%%%%%%%%%%%%%%%%%%%%
\section{Conclusions and future research lines}
\label{sec:conclusions}
%%%%%%%%%%%%%%%%%%%%%%%%%%%%%%%%%%%%%%%%%%%%%%%%%%%%%%%%%%%%%%%%

The BIRME 3D method has been adapted in this work to analyze the detection of GWs using microwave resonant cavities. Whilst the classical analysis of these cavities with numerical methods (finite elements or finite differences) provide scattering parameters or eigenvalues/eigenvectors that only allow obtaining the resonance characteristics (resonant frequency, loaded and unloaded quality factor, and the form factor), this new formulation can introduce the source, that is, the GW induced current $\vec{J}_{GW}$, and to obtain the magnitude and phase of the signal produced by the GWs at the output ports of the cavity. On the one hand, this allows one to precisely obtain the detected power or voltage responses over a wide frequency range, obviating the need for Cauchy-Lorentz approximations. On the other hand, this analysis method enables, for the first time, the acquisition of the phases of the signals in the output ports, which may be a crucial consideration when attempting to develop GW microwave-cavity detectors that operate with interferometric methods among various cavities.
If one considers shortly-lived GWs, the detection time in the receiver cannot be increased too much, and interferometry may be a powerful technique for getting better sensitivities.

This formulation has been applied to a cubic cavity with three perpendicular ports which allow for the simultaneous detection of the three degenerate modes $TE_{101}$, $TE_{011}$ and $TM_{110}$. It has been shown that, in some cases, the combination of two modes increases the range of incident angles that can be explored while maintaining high coupling levels.
This range is complete for the $XY$ plane case. For $XZ$ and $YZ$ planes, possible solutions for avoiding blind angles are introducing a second cavity (rotated with regards to the first one) or performing a rotation of the cavity inside the magnet. Both actions lead to a rotation of the external magnetic field inside the cavity. Notice also that the natural rotation of the Earth would improve this situation for signals persisting for hours. 

It has been confirmed that the coupling of the GW with the cavity modes is much more complex than the axion one. In the latter, there exists a clear dominant mode with optimal form factor, but for GWs any mode, regardless of its polarization, can get an adequate coupling, depending on the incident angle and the frequency. This motivates the determination of how the detection can be improved by combining the extracted signals from different modes. In an alternative operation mode, by detuning degenerate modes and probing other higher-order modes, a set of $N$ frequencies (one per mode) can be explored in parallel. Another key difference that we have not remarked on until now is that the tuning of the signal to a resonant mode of the cavity may happen {naturally} for GWs, without the need for a scanning strategy. This is because when black hole binaries emit GWs in this range, their orbital motion typically evolves fast enough to explore the frequencies of interest in their emitted signal in a relatively short time. In this sense, by focusing on integration times of ms, one expects the narrow resonances shown in figures \ref{h_XZ_YZ_XY_C1}-\ref{h_XZ_YZ_XY_C3} to be excited during the merger event. In the future, we plan to study the effect of considering the frequency spectrum of a GW using the formulation developed in this work.

Although this work has been done on a relatively basic rectangular-section microwave cavity, it can be readily expanded to other cavities with superior performance characteristics, such as cylindrical, spherical, or other mixed geometries like rectangular cavities with bent edges, reducing losses at the edges and so improving the quality factor.  Moreover, although the simulations show detected voltage and power levels still far from expected signals, the presented setup is a first step for GW detection for a wide range of incident angles. Further improvements can be put in place to work towards a more relevant sensitivity. First, the cavity performance indicator, $Q_0V^{5/3}\tilde\eta_{+,\times}^2$, can be improved following the main lines currently in development by the axion detection community. For instance, a better quality factor can be achieved by using {type II} superconductor materials \cite{Posen:2022tbs,Ahn:2020qyq} or dielectric materials \cite{DiVora:2022tro}, maintaining the three degenerate modes. {In fact, in our preliminary simulations, the quality factor for this cubic cavity is increased by a factor $\gtrsim 10$ when rare-earth barium copper oxide (ReBCO) superconductors are used.} Another critical aspect for the cavity design in future works will be the introduction of a tuning system that affects the same way the three degenerate modes. Remarkably, this tuning system may also be used to control the splitting of these modes, which may serve as a way to detect GWs of much lower frequencies, following the ideas\footnote{These ideas are based on cavities loaded with a mode, that is transformed into another one by the arrival of GWs. We leave the precise study of these set-ups for future works.} in \cite{Ballantini:2005am,Berlin:2023grv}. Furthermore, it would be very relevant to reduce the noise in the haloscope system using photon detection devices, such as qubits, already proposed for dark photon detection \cite{Dixit:2020ymh}, although magnetic resilience is still not developed.  Altogether, these improvements (superconductors, cavity geometry, 3D transmon detection) may boost the sensitivity of the experiment by a factor $10\thicksim100$.

\acknowledgments

We thank Camilo Garc\'ia-Cely for helpful comments and discussions as well as for his support in the calculation of the effective induced currents. We are also grateful to Valerie Domcke,  Claudio Gatti, and Nick Rodd for comments on a previous draft.

This work was performed within the RADES group; we thank our colleagues for their support.

It has been funded by MCIN/AEI/10.13039/501100011033/ and by ``ERDF A way of making Europe'', under grants PID2019-108122GB-C33, PID2022-137268NB-C53, PID2022-137268NA-C55, and PID2020-115845GB-I00. Generalitat Valenciana has also funded this work under the project ASFAE/2022/013. This article is based upon work from COST Action COSMIC WISPers CA21106, supported by COST (European Cooperation in Science and Technology).
%DB is supported by a `Ayuda Beatriz Galindo Senior' from the Spanish `Ministerio de Universidades', grant BG20/00228.
IFAE is partially funded by the CERCA program of the Generalitat de Catalunya. Diego Blas acknowledges 
the support from the Departament de Recerca i Universitats from Generalitat de Catalunya to the Grup de Recerca `Grup de F\'isica Te\`orica UAB/IFAE' (Codi: 2021 SGR 00649).

\begin{figure}[htbp]
\centering
\includegraphics[width=.9\textwidth]{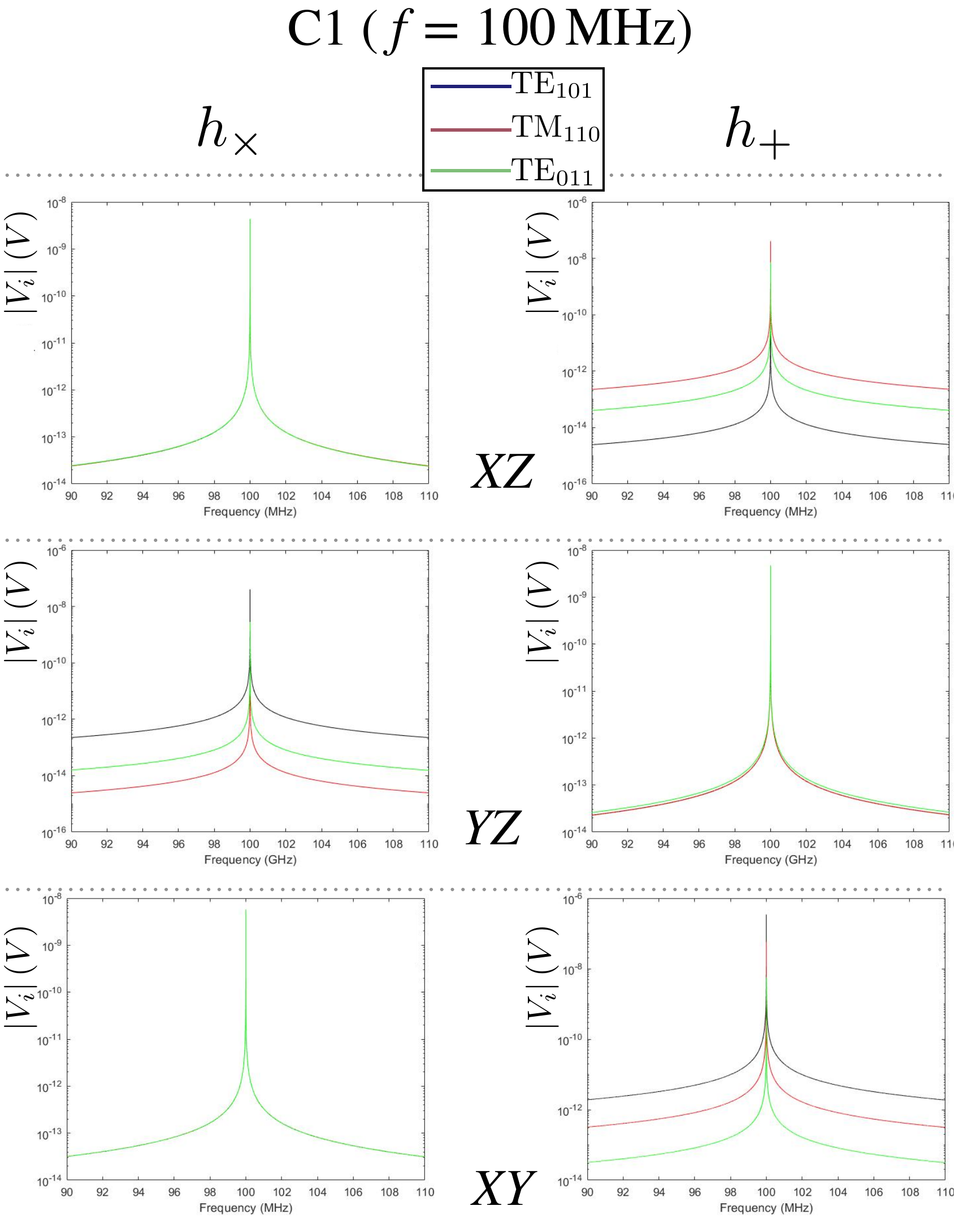}
\caption{Magnitude of the detected voltages $|v_i|$ as a function of frequency in the three coaxial probes of the cavity C1 ($f \, = \, 100$ MHz). Magnetostatic field: $B_0 = 0.6$ T; signal-to-noise ratio: $S/N=3$; temperature of the system: $T_{sys} = 1$ K; frequency detection bandwidth: $\Delta f = 5$ kHz;  detection time: $\Delta t = 1$ ms. Left: cross-polarization; Right: plus polarization. Up: GW incidence in the $XZ$ plane; Center: GW incidence in the $YZ$ plane; Down: GW incidence in the $XY$ plane.}
\label{vi_magnitude_XZ_YZ_XY_C1}
\end{figure}
\begin{figure}[htbp]
\centering
\includegraphics[width=.9\textwidth]{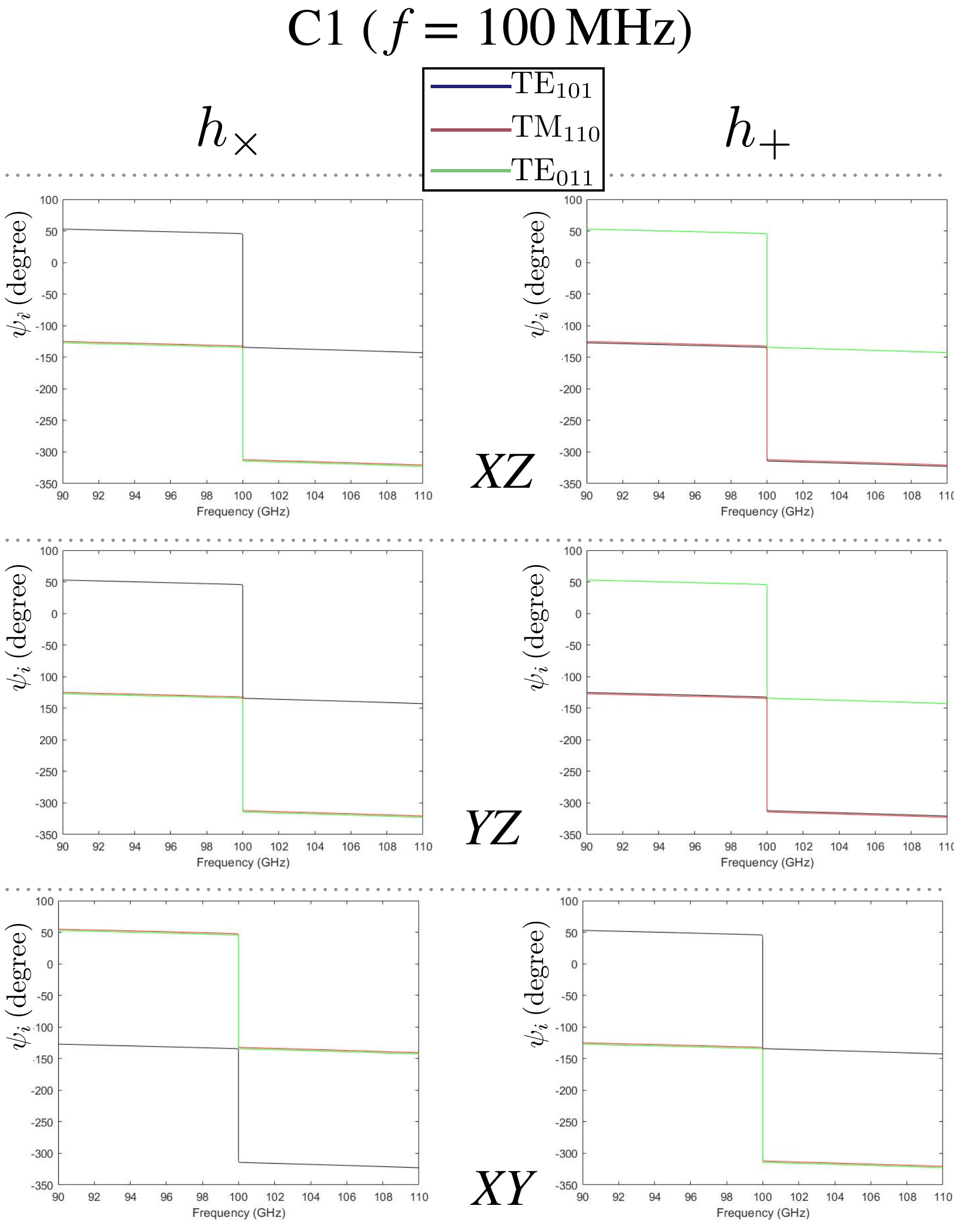}
\caption{Phase of the detected voltages $\psi_i$ as a function of frequency in the three coaxial probes of the cavity C1 ($f \, = \, 100$ MHz). Magnetostatic field: $B_0 = 0.6$ T; signal-to-noise ratio: $S/N=3$; temperature of the system: $T_{sys} = 8$ K; frequency detection bandwidth: $\Delta f = 5$ kHz;  detection time: $\Delta t = 1$ ms. Left: cross-polarization; Right: plus polarization. Up: GW incidence in the $XZ$ plane; Center: GW incidence in the $YZ$ plane; Down: GW incidence in the $XY$ plane.}
\label{vi_single_phase_XZ_YZ_XY_C2}
\end{figure}

\bibliographystyle{JHEP}

\bibliography{biblio.bib}
\end{document}